\documentclass[11pt,a4,twoside]{article}
\usepackage[dvips,a4paper,left=2.92cm,right=2.92cm,top=2.0cm,bottom=2.0cm,headsep=1em]{geometry}
\usepackage{fancyhdr}
\usepackage{titlesec}
\usepackage[latin1]{inputenc}
\usepackage{amsmath,amsfonts}
\usepackage{rotating}
\usepackage[font={small}]{caption}
\usepackage{natbib}
\usepackage{lmodern,slantsc}
\usepackage[breaklinks,
colorlinks=true, linkcolor=blue, citecolor=black, urlcolor=blue,
pdfborder={0 0 0}, pdfpagelabels]{hyperref}
\usepackage{tikz}
\usepackage{enumitem}
\def\vx{\mathbf x}

\def\va{\mathbf a}

\def\vf{\mathbf f}

\def\vn{\mathbf n}

\def\vv{\mathbf v}

\def\vA{\mathbf A}

\def\vC{\mathbf C}
\def\vD{\mathbf D}

\def\vG{\mathbf G}
\def\vH{\mathbf H}
\def\vI{\mathbf I}
\def\vL{\mathbf L}

\def\vU{\mathbf U}
\def\vV{\mathbf V}

\def\v0{\boldsymbol{0}}

\def\L{\langle}
\def\R{\rangle}
\newlength{\FigureHeight}
\newlength{\FigureHeightHalf}

\numberwithin{equation}{section}
\titleformat{\section}
{\large\bfseries}{\thetitle.}{0.5em}{}
\titleformat{\subsection}
{\normalfont\itshape\filcenter}{\normalfont\thetitle.}{0.5em}{}
\begin{document}

\title{A critical examination of the statistical symmetries admitted by the
Lundgren-Monin-Novikov hierarchy \\ of unconfined turbulence}
\author{M. Frewer$\,^1$\thanks{Email address for correspondence:
frewer.science@gmail.com}$\:\,$, G. Khujadze$\,^2$ \& H. Foysi$\,^2$\\ \\
\small $^1$ Tr\"ubnerstr. 42, 69121 Heidelberg, Germany\\
\small $^2$ Chair of Fluid Mechanics, Universit\"at Siegen, 57068
Siegen, Germany}
\date{{\small\today}}
\clearpage \maketitle \thispagestyle{empty}

\vspace{-2em}
\begin{abstract} \noindent
We present a critical examination of the recent article
by~\cite{Oberlack14.1} which proposes two new statistical
symmetries in the classical theory for turbulent hydrodynamic
flows.~We first show that both symmetries are unphysical in that
they induce inconsistencies due to violating the principle of
causality. In addition, they must get broken in order to be
consistent with all physical constraints naturally arising in the
statistical Lundgren-Monin-Novikov (LMN) description of
turbulence. As a result, we state that besides the well-known
classical symmetries of the LMN equations no new statistical
symmetries exist. Yet, aside from this particular issue, the
article by \cite{Oberlack14.1} is flawed in more than one respect,
ranging from an incomplete proof, to a self-contradicting
statement up to an incorrect claim. All these aspects will be
listed, discussed and corrected, thus obtaining a completely
opposite conclusion in our study than the article by
\cite{Oberlack14.1} is proposing.

\vspace{0.5em}\noindent{\footnotesize{\bf Keywords:} {\it
Turbulence, Lie Groups, Symmetries, Probability Density Functions,
Principle of Causality, Closure Problem, Scaling Laws, Random
Galilean Invariance,
Intermittency}}$\,$;
\end{abstract}

\section{Introduction\label{S1}}

In order to examine the article by
\href{http://journals.aps.org/pre/abstract/10.1103/PhysRevE.90.013022}
{Wac{\l}awczyk~{\it et~al.}~(2014)} as clearly as possible, we
will analyze it step by step and list all our objections according
to their relevance in~how they contribute to the articles' two
main conclusions, in that: (i) a new turbulent scaling law is
proposed which has been derived as an invariant solution from two
new statistical symmetries, and (ii) one of which, namely the new
statistical {\it scaling} symmetry, has been identified as a
measure for the intermittent flow behavior of the velocity signal.
Unfortunately, however, both these conclusions cannot be confirmed
when considering the following facts we present in this study. But
before we begin our critical examination, a brief summary of
\cite{Oberlack14.1} will be helpful: Based on the instantaneous
multi-point velocity correlation moments of order~$n+1$
\begin{align}
H_{i_{\{n+1\}}}=H_{i_{(1)}i_{(2)}\dotsc i_{(n+1)}}=\L
U_{i_{(1)}}(\vx_{(1)},t)\cdots U_{i_{(n+1)}}(\vx_{(n+1)},t)\R,\;\;
n\geq 0,\label{1-01}
\end{align}
which evolve according to the Friedmann-Keller multi-point
correlation (MPC) equations
\begin{equation}
\frac{\partial H_{i_{\{n+1\}}}}{\partial t}+\sum_{l=1}^{n+1}\left[
\frac{\partial H_{i_{\{n+2\}}[i_{(n+2)}\mapsto
k_{(l)}]}[\vx_{(n+2)}\mapsto\vx_{(l)}]}{\partial
x_{k_{(l)}}}+\frac{\partial I_{i_{\{n\}}[l]}}{\partial
x_{i_{(l)}}}-\nu\frac{\partial H_{i_{\{n+1\}}}}{\partial
x_{k_{(l)}}\partial x_{k_{(l)}}}\right]=0,\label{1-02}
\end{equation}
along with the continuity constraints for the velocity moments
\begin{equation}
\frac{\partial H_{i_{\{n+1\}}[i_{(l)}\mapsto k_{(l)}]}}{\partial
x_{k_{(l)}}}=0,\;\; \text{for $l=1,\dotsc ,n+1$},\label{1-03}
\end{equation}
and for the velocity-pressure moments
\begin{equation}
\frac{\partial I_{i_{\{n\}}[k][i_{(l)}\mapsto m_{(l)}]}}{\partial
x_{m_{(l)}}}=0,\;\; \text{for $k,l=1,\dotsc , n+1$, and $k\neq
l$},\label{1-04}
\end{equation}
where the latter moments are defined as
\begin{multline}
\; I_{i_{\{n\}}[l]}=\L U_{i_{(1)}}(\vx_{(1)},t)\cdots
U_{i_{(l-1)}}(\vx_{(l-1)},t)\cdot P(\vx_{(l)},t)\\ \cdot
U_{i_{(l+1)}}(\vx_{(l+1)},t)\cdots
U_{i_{(n+1)}}(\vx_{(n+1)},t)\R,\;\;\label{1-05}
\end{multline}
it is argued in \cite{Oberlack14.1} that next to the classical
statistical symmetries
\begin{align}
\bar{T}_1: & \;\;\; t^*=t+a_1,\;\;\; \vx^*_{(l)}=\vx_{(l)},\;\;\;
\vH^*_{\{n\}}=\vH_{\{n\}},\;\;\;
\vI^*_{\{n\}}=\vI_{\{n\}},\label{1-06}\\[0.75em]
\bar{T}_2: & \;\;\; t^*=t,\;\;\;
\vx^*_{(l)}=e^{a_2}\vx_{(l)},\;\;\; \vH^*_{\{n\}}=e^{n\cdot
a_2}\vH_{\{n\}},\;\;\;
\vI^*_{\{n\}}=e^{(n+2)a_2}\vI_{\{n\}},\label{1-07}\\[0.75em]
\bar{T}_3: & \;\;\; t^*=e^{a_3}t,\;\;\;
\vx^*_{(l)}=\vx_{(l)},\;\;\; \vH^*_{\{n\}}=e^{-n\cdot
a_3}\vH_{\{n\}},\;\;\;
\vI^*_{\{n\}}=e^{-(n+2)a_3}\vI_{\{n\}},\label{1-08}\\[0.75em]
\bar{T}_4-\bar{T}_6: &\;\;\; t^*=t,\;\;\;
\vx^*_{(l)}=\va\cdot\vx_{(l)},\;\;\;
\vH^*_{\{n\}}=\vA_{\{n\}}\otimes\vH_{\{n\}},\;\;\;
\vI^*_{\{n\}}=\vA_{\{n\}}\otimes\vI_{\{n\}},\label{1-09}\\[0.75em]
\bar{T}_7-\bar{T}_9: &\;\;\; t^*=t,\;\;\;
\vx^*_{(l)}=\vx_{(l)}+\vf(t),\;\;\;
\vH^*_{\{n\}}=\vH_{\{n\}}+\sum_{b=1}^n f^\prime_{i_{(b)}}(t)
\vH_{\{n-1\}[i_{(b)}\mapsto \emptyset]},\nonumber\\
&\;\;\; \vI^*_{\{n\}[l]}=\vI_{\{n\}[l]}-f^{\prime\prime}_\beta
x_{\beta_{(l)}}\vH_{\{n-1\}[l\mapsto \emptyset]}+ \sum_{c=1,c\neq
l}^{n+1} f^\prime_{i_{(b)}}\vI_{\{n-1\}[l][c\,\mapsto
\emptyset]},\label{1-10}\\[0.75em]
\bar{T}_{10}: &\;\;\; t^*=t,\;\;\; \vx^*_{(l)}=\vx_{(l)},\;\;\:
\vH^*_{\{n\}}=\vH_{\{n\}},\;\;\, \vI^*_{\{n\}[l]}=\vI_{\{n\}[l]}
+f_4(t)\vH_{\{n\}[i_{(l)}\mapsto \emptyset]},\label{1-11}
\end{align}
where $f$'s are free functions and
$\vA=\va_{(1)}\otimes\cdots\otimes\,\va_{(n)}$ is a n-point
concatenation of a rotation matrix~$\va\in\text{O}(3)$, the MPC
equations \eqref{1-02} admit two more statistical symmetries
\begin{align}
\bar{T}^\prime_{2_{\{n\}}}: &\;\;\; t^*=t,\;\;\;
\vx^*_{(l)}=\vx_{(l)},\;\;\;
\vH^*_{\{n\}}=\vH_{\{n\}}+\vC_{\{n\}},\;\;\;
\vI^*_{\{n\}}=\vI_{\{n\}}+\vD_{\{n\}},\label{1-12}\\[0.75em]
\bar{T}^\prime_{s}: &\;\;\; t^*=t,\;\;\;
\vx^*_{(l)}=\vx_{(l)},\;\;\;
\vH^*_{\{n\}}=e^{a_s}\vH_{\{n\}},\;\;\;
\vI^*_{\{n\}}=e^{a_s}\vI_{\{n\}},\label{1-13}
\end{align}
which, in contrast to the above classical symmetries
\eqref{1-06}-\eqref{1-11}, are {\it not} reflected by the
underlying deterministic Euler and Navier-Stokes equations. For
$n=1$ in the velocity moments $\vH_{\{n\}}$, the translation
symmetry \eqref{1-12}
\begin{align}
\bar{T}^\prime_{2_{\{l\}}}: &\;\;\; t^*=t,\;\;\;
\vx^*_{(l)}=\vx_{(l)},\;\;\;
\vH^*_{\{1\}}=\vH_{\{1\}}+\vC_{\{1\}},\nonumber\\
& \;\;\; \vH^*_{\{n\}}=\vH_{\{n\}},\;\;\;
\vI^*_{\{m\}}=\vI_{\{m\}}+\vD_{\{m\}},\;\;\; \text{for $n\geq 2$,
$m\geq 1$,}\label{1-14}
\end{align}
is furthermore identified in \cite{Oberlack14.1} as the random
Galilean group or random Galilean invariance as was first
introduced by \cite{Kraichnan64,Kraichnan65,Kraichnan68}.

It is then shown that all symmetries for the moments
\eqref{1-06}-\eqref{1-13}\footnote[2]{Except for the extended
Galilean invariance \eqref{1-10} which has to be restricted to
$\vf^{\prime\prime}(t)=0$ in order to be consistent with a
necessary solution manifold of bounded functions.} transcribe to
symmetries for the probability density functions (PDFs)
\begin{equation}
f_n=f_n(1,\dotsc
,n)=f_n(\vv_{(1)},\dotsc,\vv_{(n)};\vx_{(1)},\dotsc,\vx_{(n)},t),
\label{1-15}
\end{equation}
which themselves evolve according the Lundgren-Monin-Novikov (LMN)
equations
\begin{multline}
\left[\partial_t +\sum_{i=1}^n\vv_{(i)}\cdot\nabla_{(i)}\right]
f_n=-\sum_{j=1}^n\nabla_{\vv_{(j)}}\cdot
\Bigg[\lim_{\vx_{(n+1)}\to\vx_{(j)}} \nu\Delta_{(n+1)}\int
d\vv_{(n+1)}\vv_{(n+1)}f_{n+1}
\\
-\int d\vx_{(n+1)}d\vv_{(n+1)}\left(\nabla_{(j)}\frac{1}{4\pi\vert
\vx_{(j)}-\vx_{(n+1)}\vert}\right)(\vv_{(n+1)}\cdot\nabla_{(n+1)})^2f_{n+1}\Bigg],
\label{1-16}
\end{multline}
which, next to the continuity conditions, go along with several
natural constraints in order to guarantee a physical solution for
the PDFs $f_n$. Considered are the {\it normalization constraint}
\begin{align}
\int d\vv_{(1)}f_1(\vv_{(1)};\vx_{(1)},t) & = 1,\nonumber\\
\int d\vv_{(2)}f_2(\vv_{(1)},\vv_{(2)};\vx_{(1)},\vx_{(2)},t) & =
f_1(\vv_{(1)};\vx_{(1)},t),\label{1-17}\\
&\;\; \vdots\nonumber
\end{align}
the {\it coincidence constraint}
\begin{align}
\lim_{\vx_{(2)}\to\vx_{(1)}}f_2(1,2) &
=f_1(1)\delta(\vv_{(2)}-\vv_{(1)}),\nonumber\\
\lim_{\vx_{(3)}\to\vx_{(1)}}f_3(1,2,3) &
=f_2(1,2)\delta(\vv_{(3)}-\vv_{(1)}),\label{1-18}\\
&\;\; \vdots\nonumber
\end{align}
and the {\it separation constraint} (here shown only for the
two-point PDF)
\begin{equation}
\lim_{\vert
\vx_{(2)}-\vx_{(1)}\vert\to\infty}f_2(1,2)=f_1(1)f_1(2).\label{1-19}
\end{equation}
According to \cite{Oberlack14.1} the corresponding transcribed
symmetries to \eqref{1-12} and \eqref{1-13} admitted by the LMN
equations \eqref{1-16} are, respectively,\footnote[2]{In
\cite{Oberlack14.1} \eqref{1-20} is derived as the result ``(42)",
and \eqref{1-21} as ``(63)".}
\begin{align}
\bar{T}^\prime_{2_{\{n\}}}: &\;\;\; t^*=t,\;\;\;
\vx^*_{(n)}=\vx_{(n)},\;\;\; \vv^*_{(n)}=\vv_{(n)},\nonumber\\
&\;\;\; f_n^*=f_n+\psi(\vv_{(1)})\delta(\vv_{(1)}-\vv_{(2)})\cdots
\delta(\vv_{(1)}-\vv_{(n)}),\label{1-20}
\end{align}
and
\begin{align}
\bar{T}^\prime_{s}: &\;\;\; t^*=t,\;\;\;
\vx^*_{(n)}=\vx_{(n)},\;\;\;
\vv^*_{(n)}=\vv_{(n)},\nonumber\\
& \;\;\; f_n^*=e^{a_s}f_n+(1-e^{a_s})\cdot\delta(\vv_{(1)})\cdots
\delta(\vv_{(n)}).\quad\;\;\;\label{1-21}
\end{align}
The invariance \eqref{1-20} is called ``shape symmetry" as it
according to \cite{Oberlack14.1} only changes the shape of the
PDF, while \eqref{1-21} is called ``intermittency symmetry"  since
all transformed PDFs $f_n^*$, when globally scaled by a same
constant factor as in \eqref{1-21}, are ``functions describing
intermittent flows" (\cite{Oberlack14.1}, p.~2).

Furthermore it is recognized that if the Lie-group structure of
both transformations \eqref{1-20} and \eqref{1-21} is reduced to
the structure of a semi-group, then compatibility with the
axiomatic constraint $f_n\geq 0\Leftrightarrow f_n^*\geq 0$ is
achieved. It is asserted that although this probabilistic
interpretation of the PDFs breaks the Lie-group structure of the
transformations \eqref{1-20} and \eqref{1-21} down to a
semi-group, they nevertheless still constitute symmetries of the
LMN equations \eqref{1-16}.

Regarding the compatibility with the other three constraints
\eqref{1-17}-\eqref{1-19}, it is claimed that for the translation
symmetry \eqref{1-20} the separation constraint \eqref{1-19} has
no influence and thus can be safely ignored for all their purposes
considered in \cite{Oberlack14.1}.

\newpage
\noindent For the scaling symmetry \eqref{1-21}, however, the
separation constraint \eqref{1-19} is only compatible if the
untransformed PDF ``is itself a $\delta$ function", because then,
according to \eqref{1-21}, one obtains ``$f_n^*(1,\dotsc ,
n)=f_n(1,\dotsc , n)=\delta(\vv_{(1)})\cdots \delta(\vv_{(n)})$",
and ``hence, if in the far field $\vert
\vx_{(1)}-\vx_{(2)}\vert\rightarrow\infty$ the one-point PDF is a
$\delta$ function $f_1(2)=f_1^*(2)=\delta(\vv_{(2)})$, then the
separation property is satisfied for the scaling invariance"
(\cite{Oberlack14.1}, p.~8):
\begin{align}
\lim_{\vert \vx_{(1)}-\vx_{(2)}\vert\to\infty}f_2^*(1,2)& =
e^{a_s}f_1(1)\delta(\vv_{(2)})
+(1-e^{a_s})\delta(\vv_{(1)})\delta(\vv_{(2)})\nonumber\\
& = f_1^*(1)\delta(\vv_{(2)})=f_1^*(1)f_1^*(2).\label{1-22}
\end{align}
Regarding the coincidence constraint \eqref{1-18}, both
transformations \eqref{1-20} and \eqref{1-21} are compatible,
while for the translation symmetry \eqref{1-20} the normalization
constraint \eqref{1-17} in connection with the axiomatic
constraint $f_n\geq 0\Leftrightarrow f_n^*\geq 0$ imposes
restrictions on the scalar function $\psi=\psi(\vv)$. These
restrictions are then derived by attempting ``to find a physical
interpretation of the shape symmetry". According to
\cite{Oberlack14.1} this is achieved by constructing a PDF which
can be written as a sum of a turbulent and laminar part
$f=f_T+f_L$, for~which then, when addressing the particular case
of fully developed plane Poiseuille channel flow, the laminar part
(due to the explicit presence of the non-moving boundaries at
$y_{(k)}=\pm H$) takes the form
\begin{multline}
f_L(\vv_{(1)},\dotsc ,\vv_{(n)};\vx_{(1)},\dotsc ,\vx_{(n)},t)
=\Bigg\L\delta\Bigg[\vv_{(1)}-\vU_0^{\{\omega\}}\Bigg(1-\frac{y_{(1)}^2}{H^2}\Bigg)\Bigg]
\\ \cdots
\delta\Bigg[\vv_{(n)}-\vU_0^{\{\omega\}}\Bigg(1-\frac{y_{(n)}^2}{H^2}\Bigg)\Bigg]\Bigg\R
,\qquad\quad\label{1-23}
\end{multline}
where the constant field
$\vU_0^{\{\omega\}}=(U_0^{\{\omega\}},0,0)$ is the streamwise
velocity in the centerline and $y_{(k)}=x_{{(k)}2}$ is the
wall-normal coordinate.

It is then argued that the ``shape symmetry" \eqref{1-20} in
channel flow takes the following form
\begin{equation}
f^*_n=f_n+F(y_{(1)},\dotsc
,y_{(n)})\psi(\vv^\prime_{(1)})\delta(\vv^\prime_{(1)}-\vv^\prime_{(2)})
\cdots \delta(\vv^\prime_{(1)}-\vv^\prime_{(n)}),\label{1-24}
\end{equation}
where
\begin{equation}
F(y_{(1)},\dotsc
,y_{(n)})=\left(1-\frac{y^2_{(1)}}{H^2}\right)^{-1}\cdots
\left(1-\frac{y^2_{(n)}}{H^2}\right)^{-1},\label{1-25}
\end{equation}
and for each k
\begin{equation}
v^\prime_{(k)1}=v_{(k)1}\left(1-\frac{y^2_{(k)}}{H^2}\right)^{-1},\;\;\;
v^\prime_{(k)2}=v_{(k)2},\;\;\;
v^\prime_{(k)3}=v_{(k)3}.\label{1-26}
\end{equation}
This symmetry \eqref{1-24} then gives rise to the following
invariant translation for the moments
\begin{multline}
\L U_1(\vx_{(1)},t)\cdots U_1(\vx_{(n)},t)\R^*=\L
U_1(\vx_{(1)},t)\cdots U_1(\vx_{(n)},t)\R \\ +C_{1\dotsc 1}
\left(1-\frac{y^2_{(1)}}{H^2}\right)\cdots
\left(1-\frac{y^2_{(n)}}{H^2}\right),\quad\label{1-27}
\end{multline}
where, according to \cite{Oberlack14.1}, the coefficients
$C_{1\dotsc 1}$ are restricted within the following ranges:
\begin{equation}
-2U_{0\text{cr}}\leq C_1\leq 2U_{0\text{cr}},\;\;\;
-U^2_{0\text{cr}}\leq C_{11}\leq U^2_{0\text{cr}},\;\;\;
-2U^3_{0\text{cr}}\leq C_{111}\leq 2U^3_{0\text{cr}},\;\;\; \cdots
\label{1-28}
\end{equation}
with $U_{0\text{cr}}$ being the critical value up to which a
laminar channel flow can be realized.

As a final result in \cite{Oberlack14.1}, the invariant solution
for the mean velocity in the {\it intermittent} flow regime of a
plane channel flow is derived. By considering i) the classical
scaling of the Navier-Stokes equations $y^*=e^{k_2}y$ and $\L
U\R^*=e^{-k_2}\L U\R$, where $k_2$ is an arbitrary constant, ii)
the new scaling symmetry $\L U\R^*=e^{a_s}\L U\R$ \eqref{1-13},
and iii) the new translation symmetry of the mean velocity $\L
U\R^*=\L U\R+C_1(1-y^2/H^2)$ \eqref{1-27}, where $C_1$ is
restricted by \eqref{1-28}, the requested invariant solution gets
determined from the characteristic equation
\begin{equation}
\frac{d\L U\R}{(a_s-k_2)\L
U\R+C_1(1-y^2/H^2)}=\frac{dy}{k_2y},\label{1-29}
\end{equation}
in the specification $a_s=k_2$ as
\begin{equation}
\L U\R
=\frac{C_1}{k_2}\ln(y)+\frac{C_1}{2k_2}\left(1-\frac{y^2}{H^2}\right)
+\mathcal{C},\label{1-30}
\end{equation}
where $\mathcal{C}$ is an arbitrary integration constant. This
final result \eqref{1-30} completes the summary of
\cite{Oberlack14.1}.

Our critical examination of \cite{Oberlack14.1} unfolds as
follows: In Section \ref{S2} the violation of the classical
causality principle by the newly proposed LMN symmetries
\eqref{1-20}-\eqref{1-21} as well as by the correspondingly
induced MPC invariances \eqref{1-12}-\eqref{1-13} is discussed.
The shortcomings of the unclosed hierarchy of the MPC equations
\eqref{1-02} are presented in Section \ref{S3}. As a consequence,
the MPC invariances \eqref{1-12}-\eqref{1-13} are to be identified
only as equivalence transformations, and not as symmetry
transformations. The conceptual difference between these two kinds
of invariant transformations as well as its implications for
constructing invariant solutions is presented in Section \ref{S4}
and \ref{S5}, respectively. The non-compatibility of the new LMN
symmetries \eqref{1-20}-\eqref{1-21} with the underlying physical
constraints \eqref{1-17}-\eqref{1-19} is discussed in detail in
Section \ref{S6}. Then, the failure of the ``shape symmetry"
\eqref{1-20} for channel flow is explained in Section \ref{S7}. As
an aside, the non-connectedness of the new MPC translation
invariance \eqref{1-14} to random Galilean invariance
\citep{Kraichnan65}, as contrarily claimed in \cite{Oberlack14.1},
is shown in\linebreak Section \ref{S8}. Finally, a general
discussion about the illusory relation between intermittency and
global symmetries is given in Section~\ref{S9}.~In Appendices
\ref{SA} and \ref{SB}, the detailed proofs of our statements and
objections are presented.

\section{Violation of the causality principle\label{S2}}

At first it should be noticed that the two new multi-point
correlation (MPC) invariances \eqref{1-12} and \eqref{1-13},
which, respectively, are induced by the Lundgren-Monin-Novikov
(LMN) symmetries \eqref{1-20} and \eqref{1-21}\footnote[2]{Up to
the pressure transformations, since in the LMN approach, with its
infinite equational hierarchy \eqref{1-16}, the pressure has been
eliminated by the continuity equation, while in the MPC approach,
with its infinite equational hierarchy \eqref{1-02}, the pressure
is explicitly included.}, only act in a purely statistical manner
without having a transformational origin of {\it any} kind in the
underlying deterministic set of equations, i.e. without having a
cause in the Navier-Stokes equations them\-selves; neither as a
distinctive symmetry, nor as any ordinary variable transformation.
As we will show later in detail, such\linebreak a pro\-perty
inevitably leads to an unphysical behavior, because it are only
the {\it deterministic} equations which due to their spatially
nonlocal and temporally chaotic behavior induce the {\it
statistical} equations, and not vice versa. In other words, any
variable transformation, i.e. irrespective of whether it
represents a symmetry or not, which only acts on a purely
statistical (averaged) level without having a deterministic
(fluctuating) origin violates the classical principle of cause and
effect, since the system would experience an effect (change in
averaged dynamics) without a corresponding cause (change in
fluctuating dynamics). Careful, we do {\it not} claim here that
the new statistical symmetries \eqref{1-12} and \eqref{1-13},
which are induced by \eqref{1-20} and \eqref{1-21} respectively,
are unphysical just because they are not reflected again as {\it
symmetries} in the original deterministic (instantaneous)
Navier-Stokes equations. Such a claim would obviously be even
incorrect, because a mean field equation can have a symmetric
structure which must not exist in its underlying fluctuating
equation. Instead, we only simply claim that these new statistical
symmetries have no mechanical cause at all from which they can
originate --- one even runs into inconsistencies as soon as one
tries to establish any such cause, which in detail we will
demonstrate in a proof further below. Hence, the classical
principle of cause and effect is clearly violated by the new
statistical symmetries\footnote[2]{To note is that we only
consider here the admitted symmetries of equations, and not of
their solutions. Because, obviously, the induced statistical
system can have solutions which are not reflected by the
underlying deterministic system, although both systems admit the
same equational symmetry. For example in turbulent channel flow,
the time-translation symmetry of the Navier-Stokes equations can
induce a temporally stationary solution in the statistical system
of equations, while in the instantaneous equations not. The
classical principle of cause and effect, however, is not violated
hereby, because, obviously, the transformation itself, i.e. the
time translation itself, is an existing and well-defined
equational transformation on both the statistical and its
underlying deterministic level, in the way that the statistical
time translation symmetry emerges as an effect from the cause of
transforming the underlying deterministic field to different
times;\linebreak in clear contrast of course to the new
statistical symmetries proposed in \cite{Oberlack14.1}, which
exist without any cause at all. For a further discussion, see also
\cite{Frewer16.1}.} proposed in \cite{Oberlack14.1}.

Let us first examine our claim for the case of the two LMN
symmetries \eqref{1-20} and \eqref{1-21}. One observes that for
both symmetries only the coarse-grained probability density
function (PDF) $f_n$ gets transformed, while the sampled values
$\vv_{(n)}$ for the fine-grained instantaneous (fluctuating)
velocity field $\vU$ at point $\vx_{(n)}$ and time $t$ stay
invariant and thus unchanged. That means that since next to the
sampled values $\vv_{(n)}$ for the fine-grained velocity field
$\vU$ also, independently, the explicit time $t$ and all spatial
measurement points $\vx_{(n)}$ are left unchanged, any arbitrary
but fixed chosen deterministic system $(t,\vx,\vU)$ will therefore
be incapable to evolve (temporally as well as spatially) during
the transformation process of \eqref{1-20} and \eqref{1-21}. Only
the statistical (coarse-grained) quantities $f_n$ change, but
which again, however, uniquely emerge from the deterministic
(fine-grained) description of that system. In other words,
although the dynamical system $(t,\vx,\vU)$ stays unchanged under
both symmetry transformations \eqref{1-20} and \eqref{1-21} on its
fine-grained level, it nevertheless undergoes a {\it global}
change $f_{n}\rightarrow f^*_{n}$ on its induced coarse-grained
level, which is completely unphysical and not realized in nature
since it obviously violates the causality principle of classical
mechanics. Keep in mind that here the coarse-grained multi-point
PDF $f_n$ is defined as an ensemble average over all possible
fine-grained realizations of the flow (see e.g.
\cite{Lundgren67,Friedrich12})
\begin{equation}
\quad f_n(\vx_{(1)},\vv_{(1)};\dotsc;\vx_{(n)},\vv_{(n)};t)=
\big\L\prod_{i=1}^n\delta(\vv_{(i)}-\vU(\vx_{(i)},t))\big\R,\quad
\label{141129:1419}
\end{equation}
where $\L \cdot\R$ denotes the averaging or coarse-graining
process, and all $\vv_{(i)}$ the sampled values of the fluctuating
or fine-grained instantaneous velocity field $\vU$ at every point
$\vx_{(i)}$ and time $t$. For the first two lowest orders, i.e.
$n=1$ and $n=2$, the general expression (\ref{141129:1419})
reduces in~\cite{Oberlack14.1}~to ``(17)" and its subsequent
expression respectively.

Note that the opposite conclusion is not the rule, i.e. a change
on the fluctuating level can occur without inducing an effect on
the averaged level.~A macroscopic or coarse-grained (averaged)
observation might be insensitive to many microscopic or
fine-grained (fluctuating) details, a property of nature widely
known as universality (see e.g. \cite{Marro14}). For example, a
high-level complex coherent turbulent structure, though a
consequence of the low-level fluctuating description, does not
depend on all its details on its lowest level. The opposite again,
however, is not realized in nature, i.e., stated differently, if
the coherent structure experiences a {\it global} change on the
averaged level, e.g. in scale or in a translational shift, it
definitely must have a cause and thus must go along with a
corresponding change on the lower fluctuating level.

Apart from this unphysical transformation behavior of \eqref{1-20}
and \eqref{1-21} on the coarse-grained level, both symmetries
consequently induce inconsistencies also on the fine-grained level
which eventually can only be resolved if these two symmetries get
broken.~Because, although by construction each of these two
symmetries leave the coarse-grained LMN equations invariant, they
are nevertheless incompatible to the defining relation
(\ref{141129:1419}), which explicitly defines the transition from
the fine-grained level $\vU(\vx_{(n)},t)$ to the coarse-grained
level $f_n$. The deeper reason for this failure in
\cite{Oberlack14.1} is that the defining relation
(\ref{141129:1419}) itself was not included or incorporated into
the construction process for the symmetries \eqref{1-20} and
\eqref{1-21} of the LMN equations.

In order to reveal this breaking of invariance, it is helpful to
mentally split relation (\ref{141129:1419}) into its left-hand
side (LHS) and into its right-hand side (RHS), where the RHS then
has to be identified as the realization of the function $f_n$ on
the LHS. Now, as both transformations \eqref{1-20} and
\eqref{1-21} induce a {\it global} change on the LHS, the equality
for invariance in (\ref{141129:1419}) then demands for a
corresponding change also on its RHS. But since the sampled
velocities $\vv_{(n)}$ in \eqref{1-20} as well as \eqref{1-21}
stay invariant, i.e. unchanged, the fine-grained velocity fields
$\vU(\vx_{(n)},t)$ then have to change in order to compensate for
the global change which occurs on the LHS. But this instantly
leads to a contradiction, because if the velocity fields
$\vU(\vx_{(n)},t)$ change then also the sampled velocities
$\vv_{(n)}$ should change respectively, but according to
\eqref{1-20} and \eqref{1-21} all $\vv_{(n)}$ may {\it not}
change, ending thus in the proclaimed incompatibility.

Hence, the defining relation (\ref{141129:1419}) not only breaks
the symmetries \eqref{1-20} and \eqref{1-21} down to non-symmetry
transformations, but ultimately also degrades them down to
non-physical transformations due to violating the principle of
causality in that an arbitrary but fixed chosen deterministic
system would experience a global change on the coarse-grained
level without experiencing any change on its fine-grained level; a
process which, as already said before, simply is not realized in
nature, and thus is completely unphysical.\footnote[2]{Note again
that the opposite case, i.e.~a change on the fine-grained and {\it
no} change on the coarse-grained level, is a constant physical
realization in nature and thus {\it not} in conflict with the
causality principle. Technically this means that the kernel on the
right-hand of \eqref{141129:1419} (the fine-grained level) can
experience a transformational change which can be averaged out to
zero, thus inducing a transformational invariance on the left-hand
side of \eqref{141129:1419} (the coarse-grained level) which in
the result then stays itself unchanged.}

This physical inconsistency can also be directly observed on the
lower statistical level of the MPC invariances \eqref{1-12} and
\eqref{1-13}. For brevity we will only consider the scaling
invariance \eqref{1-13}
--- for the translation invariance \eqref{1-12} the line of argumentation
is similar. Now, to expose this inconsistency on the MPC level, we
recall that the velocity correlation function $\vH_{\{n\}}$ in
\eqref{1-13} is nonlinearly built up by $n$ multiplicative
evaluations of the single instantaneous (fluctuating) velocity
field $\vU=\vU(\vx,t)$ according to \eqref{1-01}. With this
information at hand, the following chain of reasoning instantly
emerges: Since for $n=1$ the averaged function $\vH_{\{1\}}=\L
\vU_{(1)}\R$ scales as $e^{a_s}$ for all points $\vx_{(1)}=\vx$ in
the single physical domain $\vx$, the corresponding fluctuating
quantity $\vU_{(1)}$ has to scale in the same manner, since every
averaging operator $\L,\R$ is linearly commuting with any {\it
constant} scale factor. But this implies that any product of $n$
fluctuating fields $\vU_{(1)}\otimes\cdots \otimes\vU_{(n)}$ has
to scale~as $e^{n\cdot a_s}$, which again implies that also the
corresponding averaged quantity $\vH_{\{n\}}=\L
\vU_{(1)}\otimes\cdots\otimes \vU_{(n)}\R$ then has to~scale as
$e^{n\cdot a_s}$. Hence, for {\it all} points $\vx_{(1)}=\vx$ and
{\it all} possible configurations of the instantaneous velocity
field $\vU=\vU(\vx,t)$, this chain of reasoning symbolically
reads~as \citep{Frewer14.1}:
\begin{align}
H^*_{i_{\{1\}}}=e^{a_s} H_{i_{\{1\}}} & \Rightarrow\; \big\L
U^*_{i_{(1)}}(\vx^*_{(1)})\big\R=e^{a_s}\big\L
U_{i_{(1)}}(\vx_{(1)})\big\R,
\;\; \text{for all points $\vx_{(1)}=\vx$} \nonumber\\
& \Rightarrow\; \big\L
U^*_{i_{(k)}}(\vx^*_{(k)})\big\R=e^{a_s}\big\L
U_{i_{(k)}}(\vx_{(k)})\big\R,\;\;\text{for all}\;
k\geq 1\nonumber\\
& \Rightarrow\; \big\L U^*_{i_{(k)}}(\vx^*_{(k)})\big\R=\big\L
e^{a_s} U_{i_{(k)}}(\vx_{(k)})\big\R,\;\;\text{for all possible
configurations $\vU$} \nonumber\\
& \Rightarrow\; U^*_{i_{(k)}}(\vx^*_{(k)})=e^{a_s}
U_{i_{(k)}}(\vx_{(k)})\nonumber\\
& \Rightarrow\; U^*_{i_{(1)}}(\vx^*_{(1)})\cdots
U^*_{i_{(n)}}(\vx^*_{(n)})=e^{n\cdot a_s}
U_{i_{(1)}}(\vx_{(1)})\cdots U_{i_{(n)}}(\vx_{(n)})\nonumber\\
& \Rightarrow\;\big\L U^*_{i_{(1)}}(\vx^*_{(1)})\cdots
U^*_{i_{(n)}}(\vx^*_{(n)})\big\R =\big\L e^{n\cdot a_s}
U_{i_{(1)}}(\vx_{(1)})\cdots U_{i_{(n)}}(\vx_{(n)})\big\R\nonumber\\
& \Rightarrow\;\big\L U^*_{i_{(1)}}(\vx^*_{(1)})\cdots
U^*_{i_{(n)}}(\vx^*_{(n)})\big\R=e^{n\cdot a_s}\big\L
U_{i_{(1)}}(\vx_{(1)})\cdots U_{i_{(n)}}(\vx_{(n)})\big\R\quad\qquad\quad\nonumber\\
& \Rightarrow\; H^*_{i_{\{n\}}}=e^{n\cdot a_s} H_{i_{\{n\}}}
.\label{140127:1405}
\end{align}
For a detailed explanation of this proof in all its steps, please
refer to Appendix \ref{SA}.~Conclusion (\ref{140127:1405}) clearly
de\-mon\-strates that if the 1-point function $\vH_{\{1\}}$
globally scales as $e^{a_s}$ then the $n$-point function
$\vH_{\{n\}}$ has to scale accordingly, namely as $e^{n\cdot a_s}$
and not as $e^{a_s}$~as\linebreak dictated by relation
\eqref{1-13}. Only the former scaling $e^{n\cdot a_s}$ will
guarantee for an all-over consistent relation between the
fluctuating and averaged dynamical Navier-Stokes system. In other
words, if a dynamical system experiences a {\it global}
transformational change on the averaged level then there must
exist {\it at least} one corresponding change on the fluctuating
level~\citep{Frewer16.1}. But exactly this is not the case for
\eqref{1-13} as both $\vH_{\{1\}}$ and $\vH_{\{n\}}$ scale therein
with the {\it same} global factor, for which, thus, a
corresponding fluctuating scaling cannot be derived or
constructed, meaning that the system experiences a {\it global}
change on the averaged level with no corresponding change on the
fluctuating level --- again the classical violation of cause and
effect as was already discussed before. For the remaining
velocity-pressure correlation functions $\vI_{\{n\}}$ in
\eqref{1-13} the conclusion is accordingly.

To conclude, it should be explicitly pointed out that the above
proof (\ref{140127:1405}) clearly shows that transformation
\eqref{1-13} itself, i.e. detached from any transport equations,
leads to contradictions as soon as one considers more than one
point $(n\geq 2)$. However, for $n=1$, i.e.~for~the mean velocity
$\vH_{\{1\}}=\L\vU_{(1)}\R$ itself, no contradiction exits. Only
as from $n\geq 2$ onwards, the contradiction starts, which can be
also clearly observed when comparing to DNS data, as was done e.g.
in \cite{Frewer14.1} and \cite{Frewer16.3}:\linebreak The mismatch
of the corresponding scaling laws which involve this contradictive
scaling group \eqref{1-13} gets more strong as the order of the
moments $n$ increases. Also note again that in order to perform
proof (\ref{140127:1405}) we basically used the consistency of
$n=1$ (the first four conclusions of (\ref{140127:1405})) to show
the inconsistency for all $n\geq 2$ (the remaining conclusions
of~(\ref{140127:1405})). Thus, irrelevant of whether \eqref{1-13}
represents an invariance or not, the transformation itself leads
for $n\geq 2$ to contradictions.

Hence, the new MPC invariance \eqref{1-13}, and with similar
argument also \eqref{1-12}, are both unphysical invariances as the
classical principle of causality is violated in every respect. The
physical consistency can only be restored if
$\vC_{\{n\}}=\vD_{\{n\}}=\v0$ and $a_s = 0$, i.e. if the invariant
transformations \eqref{1-12} and \eqref{1-13} get broken. In
contrast, of course, to the classical MPC symmetries
$\bar{T}_1$-$\bar{T}_{10}$ \eqref{1-06}-\eqref{1-11}, which, by
construction, do not violate causality and thus constitute
physical symmetries since they all have their origin in the
underlying deterministic Euler or Navier-Stokes equations.

\section{Unclosed hierarchy of MPC equations\label{S3}}

Although infinite in dimension, the MPC system of equations
\eqref{1-02} is not closed, not even in a formal sense. The reason
is that \eqref{1-02} always involves more unknown functions than
determining equations, as can be easily confirmed by just
explicitly counting the number of equations versus the number of
functions to be determined.\footnote[2]{A set of equations is
defined as {\it closed} if the number of equations involved is
either equal to or more than the number of dependent variables to
be solved for. In contrast, a set of equations is defined as {\it
unclosed} if the number of equations involved is less than the
number of unknown dependent variables.} For each order $n$ in the
hierarchy the total number of dynamical equations \eqref{1-02},
along with the continuity constraint equations \eqref{1-03} and
\eqref{1-04}, is always less than the total number of unknown
functions. Hereby it should be noted that by construction
\begin{align}
G_{i_{\{n+2\}}[l]}& :=H_{i_{\{n+2\}}[i_{(n+2)}\mapsto
k_{(l)}]}[\vx_{(n+2)}\mapsto\vx_{(l)}]\nonumber\\ &\;\neq
H_{i_{\{n+2\}}},\;\; \text{for all $n\geq 0$}, \label{141019:1948}
\end{align}
i.e., that this defined term $\vG_{\{n+2\}[l]}$ appearing in
\eqref{1-02} is not a $(n+2)$-point function but only a lower
dimensional $(n+1)$-point function of $(n+2)$-th moment. Although
all components of $\vG_{\{n+2\}[l]}$ (\ref{141019:1948}) can be
uniquely constructed from the higher dimensional moments
$\vH_{\{n+2\}}$ once they are known, which also can be formally
written as
\begin{equation}
\vG_{\{n+2\}[l]}=\lim_{\vx_{(n+2)}\to\vx_{(l)}}\vH_{\{n+2\}},
\label{141019:1854}
\end{equation}
the necessary inverse construction, however, fails.~Hen\-ce, since
(\ref{141019:1854}) is a non-invertible construction, i.e.~since
$\vH_{\{n+2\}}$ {\it cannot} be uniquely constructed from
$\vG_{\{n+2\}[l]}$, these latter moments are to be identified as
unclosed functions in system \eqref{1-02} as they do not {\it
directly} enter the next higher order correlation equation. In
total this just reflects the classical closure problem of
turbulence which cannot be bypassed by simply establishing the
formal connection (\ref{141019:1854}) --- for more details we
refer to \cite{Frewer14.1}.

In order to {\it formally} close this hierarchy \eqref{1-02} one
has to extend the MPC equations at each order with the lower order
moment equations for the corresponding unclosed terms
$\vG_{\{n+2\}[l]}$ (see e.g. \cite{Pope00,Davidson04}). But this
is a formidable task, as these lower-order moment equations cannot
be organized or condensed into a single hierarchy anymore as it
can be done for the MPC equations given by \eqref{1-02}. The
reason for this jump in complexity is that the above limit
(\ref{141019:1854}) is not commuting with any spatial linear
(differential or integral) operator $\vL_{\vx_{(j)}}$, i.e. since
for $j\neq i$ we have
\begin{equation}
\vL_{\vx_{(j)}}\cdot\Big(\lim_{\vx_{(i)}\to\vx_{(j)}}
\vH_{\{n+2\}}\Big)\neq
\lim_{\vx_{(i)}\to\vx_{(j)}}\Big(\vL_{\vx_{(j)}}\cdot
\vH_{\{n+2\}}\Big),\label{141219:1920}
\end{equation}
each $(n+2)$-point equation will then consequently turn into a
lower level $(n+2)$-moment equation with new unclosed terms and
with a different formal structure than given in \eqref{1-02}.~It
is due to this non-commuting property \eqref{141219:1920} that the
number of unclosed terms increases, while at the same time the
number constraints decreases \citep{Frewer14.1}. Hence, the
overall degree of unterdeterminedness of the MPC equations even
increases when trying to formally close them; and exactly in this
sense we can say that the MPC equations \eqref{1-02}-\eqref{1-04}
(as standardly used in
\cite{Oberlack10,Oberlack11.1,Oberlack13.1,Oberlack14}) are
underdetermined, in that they involve more unknowns than
equations.

Important to note in this respect is that the LMN equations
\eqref{1-16} {\it also} involve such a one-sided (non-invertible)
`lim'-connection between the lower and higher order functions,
but, in contrast to the MPC equations, the LMN equations give
something in return in that they come along with additional
physical constraints \eqref{1-17}-\eqref{1-19} in order to
constitute themselves as a {\it formally} closed system. In fact,
these additional internal constraints will restrict the infinitely
many possible (unphysical) solution manifolds of the LMN evolution
equations down to~a unique (physical) solution manifold
\citep{Frewer14.1}.~Moreover, the LMN equations~are just the
discrete version of the functional Hopf equation
\citep{Hopf52,McComb90,Shen91}, which formally represents itself
as a fully closed~equation~\citep{Monin67}.

In other words, only due to the fact that the LMN equations go
along with additional physical constraints, they constitute in
contrast to the MPC equations a {\it formally} closed system, and
thus also a more physical system, as it was already recognized
by~\cite{Ievlev70}: ``However, the equations for the probability
distributions [the LMN equations] yield a more complete and
compact statistical description of turbulence than do the usual
moment equations [the MPC or Friedman-Keller equations] and
apparently permit an easier formulation of the approximate
conditions closing the equations."

Finally note that while the transport equations for the moments
(MPC) can be uniquely determined from the evolution equations of
the PDFs (LMN) (see e.g. \cite{Monin67}), the reverse cannot be
established. In other words, the LMN equations cannot be {\it
uniquely} determined from the MPC equations, which was shown in
\cite{Ievlev70} (see also~\cite{Monin75,Frewer14.1}).~Hence, the
MPC equations are {\it not} equivalent to the LMN~equations.

\section{Equivalence versus symmetry transformations\label{S4}}

Considering the previous fact that the MPC hierarchy \eqref{1-02}
is unclosed,~it now has consequences when performing an invariance
analysis upon it. Instead of generating symmetry transformations,
only a weaker class of equivalence transformations can be
generated (see e.g.
\cite{Ovsiannikov82,Ibragimov94,Meleshko96,Ibragimov04,Bila11}).
The reason is that the MPC system of equations, although infinite
in dimension,

\pagebreak[4]\noindent is underdetermined in that it not only
involves the unknown and thus arbitrary functions
$\vG_{\{n+2\}[l]}$ (\ref{141019:1948}), but also in that it's
defined forward recursively \citep{Frewer15.1}. Hence, the
invariant MPC transformations \eqref{1-12} and \eqref{1-13} are
not to be interpreted as symmetry transformations as misleadingly
in \cite{Oberlack14.1}, but only as equivalence transformations.
The consequences for this insight will be discussed in
the~next~section.

The problem is that the quest for finding {\it symmetry}
transformations for a given set of differential equations is
fundamentally different to that for finding {\it equivalence}
transformations. The knowledge of symmetry transformations is
mainly used to construct special or general solutions of
differential equations, while equivalence transformations are used
to solve the equivalence problem for a certain class of
differential equations by group theory, that is, to find general
criteria whether two or more different differential equations are
connected by a change of variables drawn from a transformation
group.~The difference between these two kinds of invariant
transformations is defined as \citep{Frewer14.1}:

%\begin{itemize}[leftmargin=*]
\begin{itemize}
%
%\vspace{-1.0em}
\item A {\it symmetry} of a differential equation
is a transformation which maps every solution of the differential
equation to another solution of the {\it same equation}. As a
consequence a symmetry transformation leads to complete form-{\it
indifference} of the equation.~It results as an invariant
transformation if the considered equation is {\it closed}.
%
%\vspace{-1.0em}
\item An {\it equivalence transformation} for a
differential equation in a given class is a change of variables
which only maps the equation to another equation in the {same
class}. As a consequence an equivalence transformation {\it only}
leads to a weaker form-{\it invariance} of the equation. It
results as an invariant transformation either if existing
parameters of the considered equation get identified as own
independent variables, or if the considered equation itself is
{\it unclosed}.
\end{itemize}

\section{On the concept of an invariant solution in turbulence\label{S5}}

In order to understand and recognize the subtle difference between
a symmetry and an equivalence transformation in its full spectrum,
we will discuss this difference again, however, from a second,
different perspective, from the perspective of generating
invariant solutions.

First of all, one should recognize that the Lie algorithm to
generate invariant transformations for differential equations can
be equally applied in the same manner without any restrictions to
{\it under-}, {\it fully-} as well as {\it overdetermined} systems
of equations
\citep{Ovsiannikov82,Stephani89,Olver93,Ibragimov94,Andreev98,Bluman96,Meleshko05},
even if the considered system is infinite dimensional (see e.g.
\cite{Frewer15.1,Frewer15.2}). However, only for {\it fully} or
{\it overdetermined} systems these invariant\linebreak Lie
transformations are called and have the effect of {\it symmetry}
transformations, while for {\it underdetermined} systems these
invariant Lie transformations are called and have the weaker
effect of {\it equivalence} transformations.

In other words, although both a symmetry as well as an equivalence
transformation form a Lie-group which by construction leave the
considered equations invariant, the action and the consequence of
each transformation is absolutely different.~While a {\it
symmetry} transformation always maps a solution to another
solution of the {\it same equation}, an equivalence
transformation, in contrast, generally only maps a {\it possible}
solution of one underdetermined equation to a {\it possible}
solution of {\it another underdetermined equation}, where in the
latter case we assume of course that a solution of an
underdetermined equation can be somehow constructed or is somehow
given beforehand.

Now, it is clear that {\it if} for an unclosed and thus
underdetermined equation, or a set of equations, the unclosed
terms are {\it not} correlated to any existing underlying theory,
the construction of an invariant solution will only be a
particular and non-privileged solution within an infinite set of
other possible and equally privileged solutions. But, on the other
hand, if the unclosed terms are in fact correlated to an
underlying theory, either in that they\hfill underly\hfill a\hfill
specific\hfill but\hfill yet\hfill analytically\hfill not\hfill
accessible\hfill process\hfill or\hfill in\hfill that\hfill
they\hfill show

\pagebreak[4]\noindent some existing but yet unknown substructure,
the construction of an invariant {\it solution} is misleading,
and, if no prior modelling assumptions for the unclosed terms are
made, essentially even ill-defined (for more details we refer to
\cite{Frewer14.1}).

The statistical theory of turbulence, however, is exactly such a
case in which the corresponding unclosed statistical moment
equations are connected and thus correlated to an underlying
theory, namely to the deterministic Navier-Stokes theory. That is,
the MPC equations given by \eqref{1-02} are {\it not} arbitrarily
underdetermined, but underdetermined in the sense that {\it all}
unclosed terms $\vG_{\{n+2\}[l]}$ (\ref{141019:1948}) can be
physically and uniquely determined from the {\it single}
underlying but analytically non-accessible instantaneous
(fluctuating) velocity field $\vU=\vU(\vx,t)$ according to
\eqref{1-01}. Hence, no {\it real} solutions and thus also no {\it
real} invariant solutions can be determined as long as no prior
modelling procedure is invoked to close this system of equations.

If, however, within the theory of turbulence such invariant
results are nevertheless generated, they must be carefully
interpreted as only being functional relations or functional
complexes which stay invariant under the derived equivalence
group, and not as being privileged {\it solutions} of an
associated underdetermined system, as was also done e.g. in
\cite{Oberlack03,Khujadze04,Guenther05,Oberlack06,Oberlack10,She11,Oberlack13.1,Oberlack14}.
As a consequence, expression \eqref{1-30} cannot be identified as
an invariant {\it solution} to the MPC equations \eqref{1-02} as
misleadingly done by the authors in \cite{Oberlack14.1}, but only
as an invariant {\it function}, which, by construction, stays
invariant under the considered transformation groups. In this
sense function \eqref{1-30} only possibly but not necessarily may
serve as a useful turbulent scaling law. However, since there can
be only {\it one} physical realization for {\it all} moments,
which all are driven by the same single deterministic velocity
field $\vU$ according to \eqref{1-01}, and since the performed
statistical invariance analysis in \cite{Oberlack14.1} did not
appropriately involve this underlying deterministic layer of
description, the chances are extremely low that exactly this
particular invariant function \eqref{1-30} should represent the
statistically correct and thus for all correlation orders $n$
(within the hierarchy \eqref{1-02}) consistent solution to the
complex inertial scaling problem of incompressible wall-bounded
turbulence. Aside from the fact that \eqref{1-30} is based on
unphysical reasoning, the deeper reason for this negative result,
in being unable to obtain a first-principle wall-bounded turbulent
scaling law, is that currently no method (including the invariant
Lie-group method) exists to establish a profound and at the same
time an analytically accessible and correct connection between the
deterministic and the statistical description of the wall-bounded
Navier-Stokes~theory.

Important to note is that up to now only in the specific case of
homogeneous isotropic turbulence
\citep{Monin75,Davidson04,Sagaut08} all those invariant functional
complexes which are gained from equivalence scaling groups can be
further used to yield more valuable results, in particular the
explicit values for the decay rates (as done e.g. in
\cite{Oberlack02}), since one has exclusive access to additional
nonlocal invariants such as the Birkhoff-Saffman or the
Loitsyansky integral. However, for wall-bounded flows it is not
clear yet how to use or exploit such invariant functional
complexes in a meaningful way, since up to now no additional
nonlocal invariants are known.

\section{Non-compatibility with all LMN constraints\label{S6}}

From a physical point of view the main difference between the MPC
equations \eqref{1-02} and the LMN equations \eqref{1-16} is that
the latter (up to the continuity constraints of incompressibility)
come along with several additional natu\-ral (physical)
constraints. Three of them are listed by \eqref{1-17},
\eqref{1-18} and \eqref{1-19} to which any solution of
\eqref{1-16} must be consistent with in order to be attributed as
a physical solution. Hence, the symmetry transformations
\eqref{1-20} and \eqref{1-21} must be consistent with these three
constraints, otherwise physical solutions get mapped to unphysical
ones.

In sections ``D.1." and ``D.2." of \cite{Oberlack14.1} the authors
correctly show that both symmetries \eqref{1-20} and \eqref{1-21}
are fully compatible with the normalization constraint
\eqref{1-17} (up to the natural condition ``(46)" for the
arbitrary function $\psi$ in \eqref{1-20}) and the coincidence
constraint \eqref{1-18}. At the same time, however, the authors
create the wrong impression that in particular for symmetry
transformation \eqref{1-20} the separation constraint \eqref{1-19}
can actually be ignored for their purposes without consequences in
realizing that this property ``is never used in the derivation of
the equations of the LMN hierarchy", and that it's ``not satisfied
by the corresponding symmetries of the MPC equations, either"
(\cite{Oberlack14.1},~p.$\,\,$6). But, from a physical point of
view it should already be clear that the MPC equations are not the
appropriate reference to make such a conclusion since they exhibit
a fundamental disadvantage over the LMN equations when performing
an invariance analysis upon them, in particular as the MPC system
is {\it not} equivalent to the LMN system (see discussion in
Section~\ref{S3}).

On the other side, for symmetry transformation \eqref{1-21} they
misleadingly give a proof that for a {\it certain specification}
of the PDF this symmetry is compatible with the separation
constraint \eqref{1-19}. But, this proof given in \eqref{1-22} is
a {\it tautology}: By specifying the PDF $f_n$ as the spatially
independent and zero-valued $\delta$-function
$f_n=\delta(\vv_{(1)})\cdots\delta(\vv_{(n)})$, the symmetry
transformation \eqref{1-21} leads to the expression $f^*_n=f_n$,
i.e. this particular specification turns the symmetry~\eqref{1-21}
into a trivial identity transformation, which, of course in a
trivial manner, is always compatible to any thinkable constraint,
also to the separation constraint \eqref{1-19}. Obviously,
formulation \eqref{1-22} thus does not constitute a
proof-by-example that symmetry \eqref{1-21} is compatible to
constraint \eqref{1-19}, nor does it form a mathematical proof per
se since it obviously is not covering the remaining infinite set
of all other functional specifications for a possible PDF. But if,
then it can be readily recognized that the above tautological
specification of a trivial zero-valued $\delta$-function for $f_n$
is in effect the only possible PDF-specification which allows for
such a compatibility.

Hence, their interpretation on the non-compatibility of symmetry
\eqref{1-20} and their proof-by-example on the compatibility of
symmetry \eqref{1-21} regarding the separation constraint
\eqref{1-19} is misleading. The authors in \cite{Oberlack14.1}
should face the fact that both symmetry transformations
\eqref{1-20} and \eqref{1-21} are, without exceptions, {\it
incompatible} with the separation constraint \eqref{1-19}, and
that they thus violate one of the most intuitive physical
constraints of the LMN equations \eqref{1-16}\footnote[2]{Note
that although both transformations \eqref{1-20} and \eqref{1-21}
act spatially independent, the spatial limit of constraint
\eqref{1-19} nevertheless induces a separation also in the sampled
velocity values $\vv_{(n)}$, hence, \eqref{1-19} imposes an {\it
active} physical constraint on {\it both} symmetries \eqref{1-20}
and \eqref{1-21}.}: For {\it all times} every PDF solution should
show the spatial property of statistical independence when any two
points are infinitely far apart, that is, any two infinitely
distant points should not influence each other.~But exactly this
property cannot be constantly maintained when transforming the
system's variables according to \eqref{1-20} or \eqref{1-21}.~The
reason is that since both transformations are true symmetry
transformations which map solutions to new solutions of the
underlying ({\it formally} closed) equations \eqref{1-16}, and
since both at the same time are not compatible with the third
physical constraint \eqref{1-19}, we thus obtain the unwanted
effect that an initially physical solution can get mapped to an
unphysical solution in which a non-vanishing correlation between
infinitely distant points will be induced, and this for all times
since the transformations \eqref{1-20} and \eqref{1-21} do not
depend on time. This is definitely not physical and even is
strictly avoided in every PDF-modelling technique (see e.g.
\cite{Pope00,Wilczek10}).

Note that both symmetry transformations \eqref{1-20} and
\eqref{1-21} can also not be interpreted as a valid approximation
for moderate separations when the separation constraint
\eqref{1-19} is violated, as it was done by the authors in their
preceding work \citep{Staffolani14}, in particular for symmetry
\eqref{1-20}. This is due to the fact that every joint-PDF ($f_n$
for $n\geq 2$) is a spatially connected quantity, meaning that if
it does not show the correct behavior for large separations, there
is no guarantee that the same PDF will then show a realistic or
physical behavior for moderate separations. This situation can be
formally compared to a boundary value problem for a differential
equation with at least one infinite extension, in that, if the
boundary condition at infinity is not satisfied, it will effect
the solution in the whole domain. In this sense both symmetries
\eqref{1-20} and \eqref{1-21} are unphysical and only turn into
physical transformations if they also satisfy the separation
constraint, but which, however, can only be achieved if $\psi=0$
and $a_s = 0$. Hence, the separation constraint \eqref{1-19}
ultimately breaks the LMN symmetries \eqref{1-20} and \eqref{1-21}
in a non-approximative manner, and, thus, should not be used for
any further conclusions as it was incorrectly done in
\cite{Oberlack14.1} e.g.~when constructing the invariant scaling
law \eqref{1-30}.

For completeness, the reader should note that besides the three
usually mentioned LMN constraints (up to those based on the
continuity constraint), as `normalization' \eqref{1-17},
`coincidence'~\eqref{1-18} and `separation' \eqref{1-19}, there
exists a fourth constraint which unfortunately is not mentioned
anymore in the recent literature, as e.g. in \cite{Friedrich12}.
We~are talking about the additional constraint first derived in
\cite{Ievlev70} (listed therein as constraint (2.6)), and also
presented in \cite{Monin75} as constraint (19.139). Both LMN
symmetries \eqref{1-20} and \eqref{1-21} are compatible however
with this fourth constraint.

\section{On the new ``shape symmetry" in channel flow\label{S7}}

In the course of deriving the restrictions \eqref{1-28} for the
group parameters of the invariance \eqref{1-12} being induced by
the symmetry \eqref{1-20} (see second part of section ``D.1."),
the authors in \cite{Oberlack14.1} address the particular case of
a fully developed plane Poiseuille channel flow, with the result
that this so-called ``shape symmetry" \eqref{1-20} in channel flow
will take the form \eqref{1-24}. However, if we do not pose any
further restrictions on the functional structure of $\psi$, this
adapted transformation \eqref{1-24} is no longer a symmetry
transformation anymore (see Appendix \ref{SB} for proof), that is,
transformation \eqref{1-24} in its general form does not leave the
LMN equations \eqref{1-16} unchanged. Furthermore, transformation
\eqref{1-24} even induces an inconsistency on the level of the MPC
equations when transforming them according to \eqref{1-27} as an
invariant (equivalence) transformation.\footnote[2]{Note that it's
incorrect to identify \eqref{1-27} as a symmetry transformation;
at most it may only be identified as an equivalence transformation
due to the unclosed nature of the MPC equations (see discussion of
Section~\ref{S3}~\&~\ref{S4}).}

This inconsistency can be easily exposed by recognizing the
feature that the LMN equations \eqref{1-16} follow from a
derivation of the deterministic (instantaneous) incompressible
Navier-Stokes equations in which the pressure field $P=P(\vx,t)$
has been eliminated by the continuity equation (see
e.g.~\cite{Lundgren67}). For a spatially unbounded or unconfined
domain, the pressure field obeys the non-local relation (see
e.g.~\cite{McComb90})
\begin{equation}
P=\int d\vx^\prime \frac{1}{4\pi\vert\vx-\vx^\prime\vert}
\left(\nabla^\prime\otimes\nabla^\prime\right)
\cdot\left(\vU^\prime\otimes\vU^\prime\right), \label{141020:0923}
\end{equation}
i.e. the dynamics of the pressure field is dictated by the
velocity field $\vU^\prime=\vU(\vx^\prime,t)$.~When taking in
(\ref{141020:0923}) the ensemble average and transform it
according to \eqref{1-27} for the second moment of the {\it
one-point} velocity fields, we obtain the following {\it
invariant} transformation relation for the mean pressure:
\begin{align}
\L P\R & = \int d\vx^\prime \frac{1}{4\pi\vert\vx-\vx^\prime\vert}
\frac{\partial^2}{\partial x^{\prime i}\partial x^{\prime j}}\L
U^{\prime i} U^{\prime j}\R\nonumber\\
& = \int d\vx^\prime \frac{1}{4\pi\vert\vx-\vx^\prime\vert}
\frac{\partial^2}{\partial x^{\prime i}\partial x^{\prime
j}}\Bigg[\, \L U^{\prime i} U^{\prime j}\R^*
-\delta^{1i}\delta^{1j}\cdot
C_{11}\left(1-\frac{y^{\prime 2}}{H^2}\right)^2\Bigg]\nonumber\\
& = \int d\vx^\prime \frac{1}{4\pi\vert\vx-\vx^\prime\vert}
\frac{\partial^2}{\partial x^{\prime i}\partial x^{\prime j}}\L
U^{\prime i} U^{\prime j}\R^*\nonumber\\
& =\, \L P\R^*\, ,\label{141020:0936}
\end{align}
where $y^\prime=x_2^\prime=(x^{\prime 2})$ is the wall-normal
integration coordinate. Note that \eqref{1-27} is only a
translation in the field variables, while the coordinates and thus
also the integration coordinates stay invariant, i.e. $\vx^*=\vx$
and $\vx^{\prime *}=\vx^\prime$, respectively.

However, on the other hand, if we consider the ensemble-averaged
instantaneous one-point velocity equations up to first moment
\citep{Pope00,Davidson04}
\begin{equation}
\left.
\begin{aligned}
\partial_i\L U^i\R=0,\qquad\qquad\qquad\quad\\
\partial_t\L U^i\R+\partial_j \L U^i
U^j\R=-\delta^{ij}\partial_j\L P\R+\nu\Delta\L U^i\R,
\end{aligned}
~~~ \right \} \label{150302:2053}
\end{equation}
being just the continuity and the mean momentum equations
resulting from the MPC hierarchy \eqref{1-02}-\eqref{1-04} within
the one-point limit $\vx_{(n)}\rightarrow\vx_{(1)}=\vx$ for all
$n\geq 2$, and transform them according to \eqref{1-27}
\begin{equation}
\left.
\begin{aligned}
\partial_i\L U^i\R^*=0,\hspace{3.625cm}\\
\partial_t\L U^i\R^*\! +\partial_j \L U^i
U^j\R^*\! =-\delta^{ij}\partial_j\L P\R\! +\nu\Delta\L U^i\R^*\!
+\delta^{1i}\frac{2\nu C_1}{H^2},
\end{aligned}
~~~~ \right\}
\end{equation}
we obtain the following transformation
relation for the mean pressure field
\begin{equation}
\L P\R =\L P\R^*+\delta_{1k} x^k\frac{2\nu C_1}{H^2},
\label{141129:1805}
\end{equation}
if and only if we would demand invariance of the above momentum
equation as imposed by the authors in \cite{Oberlack14.1}. But
(\ref{141129:1805}) stands in contradiction to the transformation
$\L P\R=\L P\R^*$ as demanded by expression (\ref{141020:0936}).
Hence \eqref{1-27} cannot be identified as an invariant
(equivalence) transformation of the MPC equations \eqref{1-02}. If
it would, it will only lead to an inconsistency. Therefore,
\eqref{1-27} may not be used to construct an invariant function
which is linked to the LMN equations, i.e. \eqref{1-27} may not
enter expression \eqref{1-29} as it was done in
\cite{Oberlack14.1}.

The reason for this failure, in that \eqref{1-24} is not a
symmetry of the underlying LMN equations \eqref{1-16} anymore,
already lies in the equational structure of the LMN hierarchy
itself.~While system \eqref{1-16} by construction only applies for
spatially {\it unbounded} flows, the authors in
\cite{Oberlack14.1}, however, consider PDFs for {\it bounded}
flows by using particularly the expression \eqref{1-23} of a fully
developed laminar channel flow (including its boundary region at
$y=\pm H$) in order to convert their ``shape symmetry"
\eqref{1-20} into the form \eqref{1-24}.~A loss of symmetry~is
thus the consequence.~Because, when formally removing~the boundary
condition again, in letting $H\rightarrow\infty$,~will not only
restore the consistency between both transformations
(\ref{141020:0936}) and (\ref{141129:1805}) (since the latter
transformation will merge into the relation $\L P\R=\L P\R^*$ of
the former one), but will also turn transformation \eqref{1-24}
back into the symmetry \eqref{1-20} again (since for finite values
of the wall-normal coordinate the symmetry-breaking factors in
\eqref{1-24} will neutralize to $F=1$ and $\vv^\prime=\vv$).

Hence, in order to properly address PDFs for {\it bounded} flows,
the LMN hierarchy must be completely re-derived such as to
incorporate the considered boundary conditions into the
integro-differential equations.~This can be achieved for example
by using the image method for Green's functions (see e.g.
\cite{Feynman63,Jackson98}). However, this will result into a
fundamentally different functional form of equations than given by
the hierarchy \eqref{1-16}, which only applies for unbounded
flows, and to the best of our knowledge such PDF evolution
equations for bounded flows have not been derived yet. But if,
then it's also straightforward to show that the adapted ``shape
symmetry" \eqref{1-24} is not admitted as a symmetry
transformation by the bounded LMN equations for plane channel flow
either. This can be actually shown without explicitly deriving
these complex LMN equations for bounded flows.

Because, first of all, when including any specific boundaries, the
induced inconsistency presented within
\eqref{141020:0923}-\eqref{141129:1805} is maintained, which
already is a clear indication that the PDF ``shape symmetry"
\eqref{1-24} itself is not admitted as a symmetry by the
corresponding bounded LMN equations. Of course, this can also be
proven directly by investigating the invariance characteristics of
all relevant boundary terms contributing to the LMN equations.
This will be done in the second part of this investigation. Now,
for a streamwise pressure driven plane channel flow, the
instantaneous Navier-Stokes equations can be formulated as
\begin{equation}
\left.
\begin{aligned}
\partial_i U^i =0,\qquad\qquad\qquad\quad\;\;\;\\
\partial_t U^i +\partial_j \big(U^i
U^j\big)=-\delta^{ij}\partial_j P +\nu\Delta U^i +\delta^{1i}K,
\end{aligned}
~~~~ \right \} \label{151107:1531}
\end{equation}
where $K$ is a constant pressure gradient to drive the flow. These
equations \eqref{151107:1531} can then be formally solved for the
instantaneous pressure field $P$ (see e.g.
\cite{McComb90}).\footnote[2]{Note that the pressure $P$ in
\eqref{151107:1531} is an effective field with the mean field
property $\partial_x\L P\R =0$. In solving this system
\eqref{151107:1531} to obtain solution \eqref{151107:1623}, we
formally assume periodic boundary conditions for $\vU$ and $\nabla
P$ in the two homogeneous directions of the plane channel with the
corresponding boundaries lying at infinity.}\linebreak Since the
velocity field $\vU$ is restricted to be zero at the (wall)
boundary, the solution reads
\begin{align}
P=& \int_V d^3\vx^\prime\,
G(\vx,\vx^\prime)\,\left(\nabla^\prime\otimes\nabla^\prime\right)
\cdot\left(\vU^\prime\otimes\vU^\prime\right)\nonumber\\
&\qquad\qquad\qquad\quad -\nu\int_S d^2\vx^\prime\,
G(\vx,\vx^\prime)\,
\vn^\prime\cdot\Big[\left(\vn^\prime\otimes\vn^\prime\right)\cdot
\left(\nabla^\prime\otimes\nabla^\prime\right)\vU^\prime\Big],\quad\:
\label{151107:1623}
\end{align}
with the Green's function for a plane channel width of $2H>0$
given as \citep{Polyanin02}
\begin{equation}
G(\vx,\vx^\prime)\equiv
G(x-x^\prime;y,y^\prime;z-z^\prime)=\frac{1}{4\pi}\sum_{n=-\infty}^\infty
\left(\frac{1}{r_{1,n}}+\frac{1}{r_{2,n}}\right),
\label{161001:1929}
\end{equation}
where
\begin{equation}
r_{k,n}=\sqrt{(x-x^\prime)^{2\vphantom{d}}+(y+(-1)^k\hspace{0.05cm}
y^\prime+2(k-1)H-4nH)^2+(z-z^\prime)^2},\quad\; k=1,2,
\end{equation}

\noindent and, where $\vn^\prime=\vn(\vx^\prime)$ in
\eqref{151107:1623} is the unit inward normal vector at
$\vx^\prime$ on the surface (boundary) $S=\partial V$ of the flow
volume $V$, which, however, for the considered plane channel flow
just reduces to $\vn^\prime=(0,\pm 1,0)$ for the lower and upper
plate respectively.\footnote[3]{As in the fully unbounded solution
\eqref{141020:0923}, note that also the Green function
\eqref{161001:1929} in the bounded solution~\eqref{151107:1623} is
normalized as $\Delta
G(\vx,\vx^\prime)=-\delta^3(\vx-\vx^\prime)$. Further note that
the surface integral in \eqref{151107:1623} can only give a
contribution at the wall boundaries, since the surface
contributions at the (infinite far) boundaries in the two
homogeneous directions cancel each other pairwise. }

Now, by transforming the mean field equations of
\eqref{151107:1531} according to \eqref{1-27}, we will~yield
\begin{equation}
\left.
\begin{aligned}
\partial_i\L U^i\R^*=0,\hspace{4.25cm}\\
\!\!\!\!\partial_t\L U^i\R^*\! +\partial_j \L U^i U^j\R^*
=-\delta^{ij}\partial_j\L P\R\! +\nu\Delta\L U^i\R^* +\delta^{1i}K
+\delta^{1i}\frac{2\nu C_1}{H^2},
\end{aligned}
~~~\right\} \label{151101:2004}
\end{equation}
while when transforming the corresponding mean pressure $\L P\R$
of the solution \eqref{151107:1623}, we will obtain {\it again}
the invariant result $\L P\R=\L P\R^*$ as in \eqref{141020:0936},
simply due to the effect that for plane channel flow the
wall-normal vector is always orthogonal to the pure streamwise
transformation \eqref{1-27}, i.e. in due to having the particular
invariance
$\vn^\prime\cdot\L\vU^\prime\R=\vn^\prime\cdot\L\vU^\prime\R^*$.\linebreak
Hence, again, in order to achieve invariance of the above momentum
equation \eqref{151101:2004}, the mean pressure field has to
transform as \eqref{141129:1805}, but which, as before in the
unbounded case, stands in conflict with the invariant
transformation relation \eqref{141020:0936} when given through its
corresponding (now bounded) solution~\eqref{151107:1623}.

This transformational inconsistency in the equations for the
moments can also be viewed from a different perspective. Instead
of taking a fixed specific value $K$ for the constant pressure
gradient in \eqref{151107:1531}, we can alternatively treat it
also as a random but still {\it constant} quantity. For example,
we can re-formulate \eqref{151107:1531} as
\begin{equation}
\left.
\begin{aligned}
\partial_i U^i =0,\qquad\qquad\qquad\quad\quad\;\;\;\\
\partial_t U^i +\partial_j \big(U^i
U^j\big)=-\delta^{ij}\partial_j P +\nu\Delta U^i +\kappa\frac{2\nu
U_0}{H^2}\delta^{1i},
\end{aligned}
~~~~ \right \} \label{151101:1443}
\end{equation}
where $H> 0$ is the channel half-height, $U_0$ a given constant
(random) velocity scale, e.g. the velocity in the middle of the
streamwise centerline $U_0:=U^1(\vx=\boldsymbol{0},t=t_0)$
evaluated at a certain time $t=t_0$ at some fixed point in the
past $0\leq t_0<t$, and $\kappa$ a dimensionless
number\pagebreak[4] characterizing the constant pressure gradient
$K$ to drive the flow {\it independently} from the pre-set
internal velocity scale $U_0$. Now, when transforming the ensemble
averaged equations of \eqref{151101:1443} for fixed $\kappa$ over
an ensemble of different $U_0$ according to \eqref{1-27},
we~will~yield
\begin{equation}
\left.
\begin{aligned}
\partial_i\L U^i\R^*=0,\hspace{5.5cm}\\
\!\!\!\!\partial_t\L U^i\R^*\! +\partial_j \L U^i U^j\R^*\!
=-\delta^{ij}\partial_j\L P\R\! +\nu\Delta\L U^i\R^*\!
+\kappa\frac{2\nu \L U_0\R\vphantom{U^U}^*}{H^2}\delta^{1i}
+\delta^{1i}\frac{2\nu C_1}{H^2}(1-\kappa).
\end{aligned}
~~~\right\} \label{151108:1919}
\end{equation}
Hence, for $C_1\neq 0$ the mean momentum equation in
\eqref{151108:1919} is only invariant in the singular case when
$\kappa=1$, i.e. when the flow is driven by a constant streamwise
pressure gradient which is just finely tuned to the particular
(non-random) value $K=2\nu\L U_0\R/H^2$, but which, by itself,
practically only represents a hypothetical and thus non-usable
quantity, simply because its value $\L U_0\R$ is not known
beforehand. But anyhow, despite this single invariant yet
non-realizable case, the invariance in general is clearly broken
for all other pressure gradients where $\kappa\neq 1$, and, as
discussed before, will automatically induce an inconsistency in
the mean pressure transformation as soon as any invariance in
\eqref{151108:1919} is enforced.

At this stage it should be clear already that the above shown
inconsistency is actually rooted in the fact that the PDF ``shape
symmetry" \eqref{1-24} is not admitted as a symmetry by the
corresponding bounded LMN equations. Indeed, when transcribing the
pressure surface term in \eqref{151107:1623} into the LMN
formalism (see e.g.~\cite{Lundgren67}), we will additionally
obtain inside the square brackets on the right-hand side of the
original (unbounded) LMN equations \eqref{1-16} the following
surface (boundary) term\footnote[2]{In \eqref{150302:2259}, the
element $\vn_{(n+1)}$ is defined as the inward normal vector  at
point $\vx=\vx_{(n+1)}$ on the surface~$S$. For the considered
plane channel flow, however, these vectors
$\vn_{(n+1)}=\vn(\vx_{(n+1)})\equiv\vn$ are constants for
all~$n\geq 1$, in only taking the two values $\vn=(0,\pm1,0)$; see
explanation in previous footnote.}
\begin{equation}
\nu\int_S
d^2\vx_{(n+1)}\left(\nabla_{(j)}\,G(\vx_{(j)},\vx_{(n+1)})\right)\vn_{(n+1)}\cdot\Big(\vn_{(n+1)}
\cdot\nabla_{(n+1)}\Big)^2\int d\vv_{(n+1)}\vv_{(n+1)}f_{n+1},
\label{150302:2259}
\end{equation}
and, as this term will be the only viscosity-dependent
modification in the LMN equations \eqref{1-16} for unbounded
flows, it will extend the viscous part in \eqref{141021:1121} of
our proof accordingly. But, since the above term in the square
bracket already vanishes on the lowest level of order $n=1$ when
specifying $f_2=\widehat{f}_2$ \eqref{141021:0932}, i.e. since we
do not get any contributions from the above surface term
\eqref{150302:2259} for plane channel flow, all argumentations and
conclusions of the proof as given in Appendix \ref{SB} remain
valid, also for the LMN equations of bounded flows, in particular
for plane channel flow. Hence, the adapted ``shape symmetry"
\eqref{1-24} is not a symmetry transformation of the LMN
equations, neither for unbounded flows nor for the bounded plane
channel flow. The consequence of this expected result: Since the
transformation \eqref{1-24} is {\it not} a symmetry of the
underlying statistical system, it thus may {\it not} be used to
generate invariant functions, neither on the level of the PDFs
themselves, nor on the lower level of the induced moments. In the
latter case, as it was shown before, the transformation
\eqref{1-24} even induces a non-removable inconsistency in the
transformation behavior of the mean pressure field. Hence,
identifying the transformation \eqref{1-27} as the induced
symmetry from \eqref{1-24} to then generate the invariant function
\eqref{1-30}, as was done in \cite{Oberlack14.1}, is therefore
wrong and will only produce inconsistent results.

\section{The non-connectedness to random Galilean invariance\label{S8}}

The authors in \cite{Oberlack14.1} identify their statistical
transformation group \eqref{1-14} as the random Galilean group,
first introduced by \cite{Kraichnan65}. However, this
identification is not correct; the translation group \eqref{1-14}
in \cite{Oberlack14.1} is not in conformance with the random
Galilean group as defined by Kraichnan. The simple reason is that
the random Galilean group acts on the fine-grained (fluctuating)
level, while the statistical translation group \eqref{1-14}
exclusively only acts on the coarse-grained (averaged) level
without having a cause on the lower fluctuating level; a property
which itself is unphysical since it obviously violates the
classical principle of
causality~(see~Section~\ref{S2}).\pagebreak[4]

To prove this statement of non-conformance, it is necessary to
define the random Galilean group (see also \cite{McComb14} for a
recent introduction into this concept).~It emerges from the usual
deterministic Galilean transformation, which, up to a shift in
time and static rotations, i.e. only in the form of a
time-invariant boost, is given by
\begin{align}
\negthickspace\negthickspace\mathsf{G}: & \;\; t^*=t,\;\;
\vx^*=\vx+\vV\cdot t, \;\; \vU^*=\vU+\vV,\;\; P^*=P,
\label{141020:0949}
\end{align}
where $\vV$ is the velocity shift constant in time and space.
Since $\vU$ represents the instantaneous and thus fluctuating
velocity field, one has two ways now in how to turn~$\mathsf{G}$
into a physical statistical transformation.

{\it Method No.1:} If one defines $\vV$ as a non-fluctuating, i.e.
as a non-random constant variable with a statistically sharp
ensemble average $\L \vV\R =\vV$, then, taking the ensemble
average of (\ref{141020:0949}) will give the classical Galilean
transformation in statistical form
\begin{equation}
\L\mathsf{G}\R_\mathsf{1}:\;\;\; t^*=t,\;\;\; \vx^*=\vx+\vV\cdot
t, \;\;\; \L\vU\R^*=\L\vU\R+\vV,\;\;\; \L P\R^*=P,
\end{equation}
which coincides with the classical statistical symmetry
\eqref{1-10} in \cite{Oberlack14.1} when choosing $\vf(t)=\vV\cdot
t$.

{\it Method No.2:} If one defines $\vV$ as a fluctuating, i.e. as
random constant variable with ensemble average $\L\vV\R$, then,
taking again the ensemble average of (\ref{141020:0949}) will give
the random Galilean transformation
\begin{equation}
\L\mathsf{G}\R_\mathsf{2}: \;\;\; t^*=t,\;\;\; \vx^*=\vx+\vV\cdot
t, \;\;\; \L\vU\R^*=\L\vU\R+\L\vV\R,\;\;\; \L
P\R^*=P,\label{141020:0953}
\end{equation}
where \cite{Kraichnan65} chose $\vV$ with a Gaussian distribution
having zero mean $\L\vV\R=\v0$. But, also for any other
distribution with a non-zero mean $\L\vV\R\neq\v0$, the random
Galilean transformation (\ref{141020:0953}) is not connected in
any form to transformation \eqref{1-14} as claimed
in~\cite{Oberlack14.1}. Firstly, the spatial coordinates in
(\ref{141020:0953}) get transformed and do {\it not} stay
unchanged as in \eqref{1-14} (even in the Kraichnan specification
$\L\vV\R=\v0$, the spatial coordinates get transformed), and
secondly, the transformation rule for all $n$-point velocity
correlations beyond the mean velocity, i.e. for all $n\geq 2$, do
not transform invariantly as given by \eqref{1-14}, if we
consistently identify the non-vanishing and {\it constant} shift
$\vC_{\{1\}}$ of \eqref{1-14} as the {\it independent} shift
$\L\vV\R=\vC_{\{1\}}\neq 0$ of the random Galilean transformation
(\ref{141020:0953}). Because, by already considering the
transformation rule for the 2-point velocity correlation
$\vH_{\{2\}}$ based on the underlying instantaneous Galilean
transformation~(\ref{141020:0949})
\begin{align}
\vH^*_{\{2\}} &=\L \vU^*_{(1)}\otimes\vU^*_{(2)}\R=\L
(\vU_{(1)}+\vV) \otimes (\vU_{(2)}+\vV)\R\nonumber\\
&= \vH_{\{2\}}+
\Big(\vH_{\{1\}}^{(1)}+\vH_{\{1\}}^{(2)}\Big)
\otimes\vC_{\{1\}}+\L\vV\otimes\vV\R,\nonumber\\
&\neq \vH_{\{2\}},
\end{align}
one manifestly recognizes that transformation \eqref{1-14} is not
compatible with the random Galilean transformation as proposed by
\cite{Kraichnan65}, not even in a more general sense which goes
beyond a Gaussian distribution with zero mean.

\section{On capturing intermittency with global symmetries\label{S9}}

In our opinion, the method of Lie-groups, when used~as a
constructive method to generate scaling laws in particular from
{\it global} symmetry transformations, is not the appropriate
method to capture the complex spatiotemporal phenomenon of
intermittency in dynamical systems, irrespective of whether
internal (small scale) or external (large scale) intermittency is
considered. Intermittency is generally characterized by sporadic
bursts in specific physical quantities with non-Gaussian
probabilities, and, in general, is itself a property which rather
breaks than restores symmetries, not only on the fine-grained
(fluctuating) level but also on the coarse-grained (averaged)
level \citep{Tsinober13,Saint13,Brading03,Kurths95,Frisch85}.

A prominent historical example is the failure of Kolmogorov's
K41-theory \citep{Kolmogorov41.1,Kolmogorov41.2,Kolmogorov41.3},
which has been found to be increasingly inaccurate for higher
order statistics (see e.g. \cite{Frisch95,Biferale03}). Instead of
statistically restoring the deterministic scaling symmetry of the
Navier-Stokes system \citep{Fushchich93,Frisch95,Andreev98}, an
induced turbulent flow will always statistically evolve such by
showing the property of anomalous scaling and the breaking of
global self-similarity, which both interdependently can be
attributed to the complex property of intermittency
\citep{Frisch85,Frisch95,Biferale03}.

Although this example only addresses the effect of global symmetry
breaking in the process of internal intermittency, it nevertheless
serves as a representative example for any type of intermittency.
In our opinion, this effect of global symmetry breaking will be
even more pronounced for external intermittency than for internal
intermittency, which consequently will have a strong negative
impact on the overall performance of the so-called global
``intermittency symmetry" \eqref{1-21} in \cite{Oberlack14.1},
which again aims at describing external intermittent flows. The
reason for this pronounced effect is that in contrast to the
``more simpler" process of internal intermittency, which more or
less can be regarded as a {\it universal} process acting
independently from boundary conditions and which thus simply
manifests itself in {\it all} turbulent flows, the ``more complex"
external intermittency is a {\it non-universal} process which
predominately results from the interaction between the flow and
externally imposed boundary conditions. And since it's well-known
that boundary conditions are generally obstructive to global
symmetries, the mechanism of symmetry breaking will thus be more
strongly favored for external (non-universal large scale)
intermittency than it already is for internal (universal small
scale) intermittency.

The point is that even if we would only consider a homogeneous
isotropic turbulent flow (which is a highly idealized flow), the
statistical solutions, in particular the higher order
correlations, are by far more complicated than we currently can
imagine and that it's actually unrealistic to believe that this
complicated behavior can be captured by some {\it global} scaling
symmetries. The complexity even increases when considering
wall-bounded flows. Hence, proposing \eqref{1-30} as a ``solution"
to the intermittent scaling behavior of the wall-bounded plane
channel flow as done by the authors in \cite{Oberlack14.1}, it's
highly questionable if this is really the case when \eqref{1-30}
is matched to numerical or physical experiments, and, furthermore,
whether \eqref{1-30} is really consistent also to all higher order
moments when trying to involve them accordingly.

Despite the fact that this scaling law \eqref{1-30} is based on
two unphysical invariant transformations \eqref{1-12} and
\eqref{1-13} violating the fundamental principle of causality of
classical mechanics (as shown in Section \ref{S2}), and besides
the fact that they respectively emerge from two symmetries
\eqref{1-20} and \eqref{1-21} which must get broken in order to be
compatible with all\linebreak constraints of the underlying
dynamical equations (as shown in Section \ref{S6}), making thus
\eqref{1-30} as a scaling law totally unreliable, the further and
more general problem is that the\linebreak deterministic
Navier-Stokes theory itself, unfortunately, only allows for
spatially {\it global} symmetries and not for spatially {\it local
symmetries}$\,$\footnote[2]{A {\it global} symmetry of a physical
system is associated with a {\it finite} dimensional Lie-group in
which all group parameters are strict constants. On the other
hand, a {\it local} symmetry is associated with an {\it infinite}
dimensional Lie-group in which at least one group parameter is not
constant, e.g. by showing a space-time coordinate dependence of
the considered system. Most physical processes, however, only
admit spatially global symmetries since the requirement for a
spatially local symmetry is too restrictive.~But, linear quantum
theory is such a prominent example which admits a spatially local
scaling (gauge) symmetry.}: All symmetries of the deterministic
Euler and Navier-Stokes equations listed in
\eqref{1-06}-\eqref{1-11} are only {\it spatially global}
symmetries. No other, more general symmetries for this theory
exist or are known yet. And it's exactly due to this fact, that a
Lie-group based symmetry analysis for the unclosed statistical
Navier-Stokes theory didn't achieve the same great breakthrough as
it did, e.g., for the theory of quantum fields (see
e.g.~\cite{Weinberg00,Penrose05}), which is based on a spatially
local symmetry, the local gauge symmetry which successfully
predicts the unknown functional structure of the interacting
fields between the various elementary~particles.

\appendix
\titleformat{\section}
{\large\bfseries}{Appendix \thetitle.}{0.5em}{}
\numberwithin{equation}{section}

\section{Comments to Proof No.1\label{SA}}

Taken one-to-one from \cite{Frewer14.1}, the proof
(\ref{140127:1405}) reads\footnote[2]{For a better readability we
will drop in this section the redundant bracket-notation, i.e.
instead of~$\vH_{\{n\}}$ and $\vx_{(n)}$ we will write $\vH_n$ and
$\vx_n$ respectively.}:
\begin{align}
\!\!\!\! \vH^*_1=e^{a_s} \vH_1 \; & \Rightarrow\;
\big\L\vU^*(\vx^*_1,t^*)\big\R=e^{a_s}\big\L
\vU(\vx_1,t)\big\R,\;\text{for all points
$\vx_1=\vx$}\label{35}\\
& \Rightarrow\; \big\L\vU^*(\vx^*_k,t^*)\big\R=e^{a_s}\big\L
\vU(\vx_k,t)\big\R,\;\text{for all}\; k\geq 1\label{36}\\
& \Rightarrow\; \big\L\vU^*(\vx^*_k,t^*)\big\R=\big\L e^{a_s}
\vU(\vx_k,t)\big\R,\;\text{for all possible configurations $\vU$}\label{37} \\
& \Rightarrow\;\vU^*(\vx^*_k,t^*)=e^{a_s}
\vU(\vx_k,t)\label{38}\\
& \Rightarrow\;\vU^*(\vx^*_1,t^*)\otimes\cdots
\otimes\vU^*(\vx^*_n,t^*)=e^{n\cdot a_s}
\vU(\vx_1,t)\otimes\cdots \otimes\vU(\vx_n,t)\label{39}\\
& \Rightarrow\;\big\L\vU^*(\vx^*_1,t^*)\otimes\cdots
\otimes\vU^*(\vx^*_n,t^*)\big\R=\big\L e^{n\cdot a_s}
\vU(\vx_1,t)\otimes\cdots \otimes\vU(\vx_n,t)\big\R\label{40}\\
& \Rightarrow\;\big\L\vU^*(\vx^*_1,t^*)\otimes\cdots
\otimes\vU^*(\vx^*_n,t^*)\big\R=e^{n\cdot a_s}\big\L
\vU(\vx_1,t)\otimes\cdots \otimes\vU(\vx_n,t)\big\R\label{41}\\
& \Rightarrow\; \vH^*_n=e^{n\cdot a_s} \vH_n .\label{42}
\end{align}

\subsection{Comment No.1}

In \eqref{35} we identify the expression $\vH^*_1=\L\vU_1\R^*$ as
$\L \vU^*_1\R := \big\L \vU^*(\vx^*_1,t^*)\big\R$. This conclusion
is based on the simple fact that the transformation of $\vH^*_1$
is a trivial one, in which all values of $\vH_1$ just get globally
scaled by a constant factor $e^{a_s}$.

In the {\it general case}, however, a careful distinction must be
made between the two transformed expressions $\L\vU\R^*$ and
$\L\vU^*\R$, since the former directly refers to the transformed
mean velocity field while the latter refers to the transformed
instantaneous (fluctuating) velocity field which is then averaged,
and thus, {\it in general}, is mathematically distinct from the
former expression. But here we are not considering the case of
such a {\it general} variable (point) transformation
\begin{gather}
t^*=t^*\,(t,\vx_k,\vH_l),\quad \vx^*_n=\vx^*_n
(t,\vx_k,\vH_l),\quad \vH^*_n=\vH^*_n (t,\vx_k,\vH_l),\quad
(k,l)\leq n ,\label{62}
\end{gather}
between the independent variables $(t,\vx_n)$ and the dependent
variables $\vH_n$, but only, as given by \eqref{1-13}, the far
more simpler {\it specific case} of a globally uniform scaling in
the dependent variables
\begin{equation}
t^*=t,\quad \vx^*_n=\vx_n, \quad \vH^*_n=e^{a_s}\vH_n,\label{63}
\end{equation}
which, when written for example for the one-point moment at
$\vx_1=\vx$
\begin{equation}
t^*=t,\quad \vx^*=\vx, \quad \L\vU\R^*=e^{a_s}\L\vU\R ,\label{64}
\end{equation}
acts as a trivial subset of \eqref{62}. Note that in the following
we only investigate the mathematical property of the
transformation \eqref{64} {\it itself}, i.e.~whether it
additionally represents an equational invariance or not is
irrelevant. In other words, we will investigate \eqref{64} very
generally, solely as a transformation of variables detached from
any underlying transport equations.

Now, it is straightforward to recognize that particularly in this
trivial case \eqref{64}, the two above mentioned transformed
one-point expressions $\vH^*_1=\L\vU\R^*$ and $\L\vU^*\R$ are
identical
\begin{equation}
\L\vU\R^*\equiv\L\vU^*\R .\label{65}
\end{equation}
This conclusion is based on the following argument, in that we can
write
\begin{align}
\vH^*_1=e^{a_s} \vH_1 \;\; \Leftrightarrow \;\;
\L\vU\R^* & =e^{a_s} \L \vU\R\label{d43}\\
& =\L e^{a_s}\vU\R \label{d44}\\
&\underset{\text{def.}}{=:}\L \widehat{\vU}\R,\label{d45}
\end{align}
due to the fact that any constant factor as $e^{a_s}$ commutes
with every averaging operator~$\L,\R$. Hence one is able to define
a unique transformation relation $\vU\rightarrow \widehat{\vU}$ on
the instantaneous level having the {\it same} transformational
structure
\begin{equation}
\widehat{\vU}=e^{a_s}\vU,\label{66}
\end{equation}
as its averaged value given in \eqref{64}, namely a simple
multiplication of a constant factor $e^{a_s}$ on some field
values\footnote[2]{Note that if \eqref{64} would be additionally
admitted as a symmetry of some mean field transport equations,
then we may {\it not} conclude that \eqref{66} is a symmetry, too,
of the underlying instantaneous (fluctuating) equations. Because,
on the mean field level one can have a symmetric structure which
on the fluctuating level must not exist.}. This, then, uniquely
allows us to identify
\begin{equation}
\widehat{\vU}=\vU^*.\label{140921:1033}
\end{equation}
In other words, since the symbol $\widehat{\vU}$ on the left-hand
side of \eqref{66} is defined by the mathematical operation on the
right-hand side (a simple multiplication of a constant factor
$e^{a_s}$), and since this mathematical operation is exactly
identical to the right-hand side of the initial transformation
\eqref{64}, one can therefore uniquely identify the transformed
symbol on the left-hand side of \eqref{66} with the same
transformation symbol as it's used on the left-hand side of
\eqref{64}, i.e. $\widehat{\phantom{.}}\,=\!\phantom{.}^\ast$.

Again, the reason is that \eqref{64} and \eqref{66} show exactly
the same transformation structure on their right-hand sides,
namely a simple multiplication of a constant factor $e^{a_s}$ on
some field values, which then define their left-hand sides. But
since we are dealing here with the same transformational process
in \eqref{66} as in \eqref{64}, we should also explicitly display
it, namely by using $\vU^*$ and not $\widehat{\vU}$, which would
only unnecessarily overload the notation. Exactly this fact was
implicitly assumed when writing the first conclusion in
(\ref{140127:1405}).

But, as soon as we would consider a more complicated
transformation than \eqref{64}, as for example
\begin{equation}
t^*=t,\quad \vx^*=\vx, \quad \L\vU\R^*=e^{a_s(\vx)}\L\vU\R
,\label{67}
\end{equation}
where, instead of a globally constant scaling exponent $a_s$, we
now would have a {\it local} scaling exponent $a_s(\vx)$ which
explicitly depends on the spatial coordinates, the identification
\eqref{65}, of course, generally no longer holds and becomes
invalid. The reason simply is that in contrast to \eqref{64} the
scaling factor in \eqref{67} is no longer a global constant
anymore which can commute with every averaging operator $\L,\R$.
In other words, since generally
\begin{equation}
\L\vU\R^*=e^{a_s(\vx)}\L\vU\R\neq \L e^{a_s(\vx)}\vU\R, \label{68}
\end{equation}
we are no longer able to define a corresponding transformation
relation $\vU\rightarrow \vU^*$ on the instantaneous level which
has the {\it same} transformational structure as its averaged
value \eqref{67}. On the contrary, its {\it real} corresponding
transformation rule $\vU\rightarrow \vU^{**}$ will rather show a
far more complex functional structure than given by \eqref{67},
which in the first instance also cannot be mathematically
determined in a straightforward manner.

Hence, since the situation of proof (\ref{140127:1405}) is not
dealing with a complex situation like \eqref{67}, but only with a
trivial one as \eqref{64}, the notation used throughout
(\ref{140127:1405}) is correct and not misleading.

Note that already from an intuitive point of view the
identification \eqref{65} must be valid if we consider a simple
transformation as \eqref{64}. Because, since in \eqref{64} only
the field values and not the coordinates get transformed we can
perform the following thought experiment: Imagine we have an
ensemble of DNS results for the instantaneous velocity field $\vU$
(hereby it is irrelevant from which specific equations this data
set was numerically generated). From the field $\vU$ we now
construct the mean field $\L \vU\R$ (either as an ensemble average
over a set of different $\vU$, or, if we have a statistically
homogenous direction, over an integral of a single $\vU$ in this
direction). Thus we then obtain all functional values
of~$\L\vU\R$, which we now collectively multiply with a same
constant factor, say by $e^{a_s}=2$, to get the new transformed
values $\L\vU\R^*$ of \eqref{64}.

Now, the critical question: How should the underlying DNS data for
the instantaneous field $\vU$ be transformed in order to generate
with the same corresponding averaging process the just previously
constructed values $\L\vU\R^*\,$? The intuitive and correct answer
is that all DNS data must be {\it coherently} multiplied with the
same factor $e^{a_s}=2$. Only then will the new transformed data
$\vU^*$, if it emerges from the operation $\vU^*:=2\cdot \vU$ and
if it's correspondingly averaged to $\L \vU^*\R$, give the ability
to reconstruct the functional values $\L\vU\R^*$. Since there is
no other option, we hence obtain within this process the unique
result $\L\vU^*\R=\L\vU\R^*$. Of course, this reasoning is only
valid for a global (coherent) transformation as given by
\eqref{64}; for a more complicated (local) transformation as
\eqref{67} this reasoning no longer holds.

\subsection{Comment No.2}

The conclusion \eqref{38} is based on the relation \eqref{37}
which goes along with the explicit comment that this relation, by
definition, must hold for {\it all} possible configurations or
functional realizations of the fluctuating velocity field $\vU$,
and not only for any certain functional specification
$\vU=\vU_0(\vx,t)$. In this case, of course, conclusion \eqref{38}
would {\it not} be correct, because for a certain specification
$\vU=\vU_0$, we generally have the situation that $\L\vU^*_0\R=\L
\vU_0\R$ although $\vU^*_0\neq \vU_0$.

Now, for the reason that \eqref{37} must hold for {\it all}
possible configurations $\vU$, it is important to recognize that
\eqref{37} is not an equation to be solved for, but that it
represents a definition. And exactly this is the argument when
going from the third \eqref{37} to fourth line \eqref{38}. The
third line
\begin{equation}
\L \vU^*(\vx^*_k,t^*)\R= \L e^{a_s} \vU(\vx_k,t)\R,\label{88}
\end{equation}
does not stand for an equation but for a definition (as it's the
case for any variable transformation in mathematics)
\begin{equation}
\L \vU^*(\vx^*_k,t^*)\R \,\underset{\text{def.}}{:=}\,\L e^{a_s}
\vU(\vx_k,t)\R,\label{89}
\end{equation}
since the left-hand side (transformed side) is defined by the
mathematical expression and operation given on the right-hand
side. Now, since the right-hand side by definition must hold {\it
for all} possible (functional) configurations of the instantaneous
velocity field $\vU$, and since both functions $e^{a_s}\vU$ and
$\vU^*$ undergo the {\it same} operation of averaging, we thus can
only conclude that both functions themselves must be identical. In
other words, definition \eqref{89} implies the definition
\begin{equation}
\vU^*(\vx^*_k,t^*) \,\underset{\text{def.}}{:=}\, e^{a_s}
\vU(\vx_k,t),\label{90}
\end{equation}
in that the transformed instantaneous velocity field $\vU^*$ is
defined by the expression $e^{a_s}\vU$. This then gives the fourth
line \eqref{38}.

Two things should be noted here. Firstly, the above conclusion is
similar to the arguments which are standardly used in fluid
mechanics when deriving a differential conservation law from its
corresponding integral version. The similarity is given in so far
as the argument for the validity of the integral conservation law
is also based on the requirement~{\it ``for~all"}, however here,
for {\it all} possible volumes or surfaces. Hence the integral
operator from the integral conservation law can be dropped and the
integrand itself is identified as the corresponding conservation
law on the differential level.

Secondly, according to the arguments given in Comment No.1, we are
not obliged to make the direct conclusion \eqref{38} from relation
\eqref{37}. Conclusion \eqref{38} can already be directly obtained
from \eqref{35} by considering the result (\ref{140921:1033}),
i.e. we can directly conclude that
\begin{equation}
\vH^*_1=e^{a_s}\vH_1 \;\;\Rightarrow\;\; \vU^*(\vx^*_k,t^*)=
e^{a_s} \vU(\vx_k,t),\label{56}
\end{equation}
which just explicitly expresses the fact again that result
(\ref{140921:1033}) was uniquely obtained (for all $\vx_k$ within
the physical space $\vx$) as the induced transformation rule
$\vU\rightarrow\widehat{\vU}=\vU^*$ (the cause on the fluctuating
level) from the transformation rule of the mean velocity
$\vH_1\rightarrow\vH^*_1$ (the effect on the averaged level).

\subsection{Comment No.3}

In \eqref{42} we identify the transformed $n$-point velocity
correlation function $\vH^*_n$ as the transformed expression
$\big\L\vU^*(\vx^*_1,t^*)\otimes\cdots
\otimes\vU^*(\vx^*_n,t^*)\big\R$. This is just the obvious
consequence in knowing the fact that only from the transformed
velocity field $\vU^*$, as it is defined in (\ref{140921:1033})
and thus in \eqref{38}, {\it all} transformed correlation
functions $\vH^*_n$ can be uniquely defined and constructed
without inducing contradictions and without violating the
principle of causality. In other words, making the conclusion
\begin{equation}
\vH^*_1=e^{a_s}\vH_1 \;\;\Rightarrow\;\;
\vH^{*\,\prime}_n=e^{n\cdot a_s}\vH_n,\label{57}
\end{equation}
where $\vH^{*\,\prime}_n$ represents the mean product of $n$
spatial coordinate evaluations of the transformed and for all
points $\vx_n$ unique instantaneous velocity field
$\vU^*=\vU^*(\vx^*,t^*)$, and in which it then gets identified as
the transformed $n$-point correlation function $\vH^*_n$, as done
in \eqref{42},
\begin{equation}
\vH^*_n:=\vH^{*\,\prime}_n=\big\L\vU^* (\vx^*_1,t^*)\otimes\cdots
\otimes\vU^*(\vx^*_n,t^*)\big\R , \label{59}
\end{equation}
is the {\it only} possible conclusion without running into any
contractions and without violating the principle of cause and
effect.

\subsection{Comment No.4}

Note that already from an intuitive point of view only the
conclusion
\begin{equation}
\vH^*_1=e^{a_s}\vH_1 \;\;\Rightarrow\;\; \vH^*_n=e^{n\cdot
a_s}\vH_n,\label{60}
\end{equation}
given through \eqref{35}-\eqref{42}, makes physically sense, while
the conclusion induced by \eqref{1-13}, or (\ref{63})
\begin{equation}
\vH^*_1=e^{a_s}\vH_1 \;\;\Rightarrow\;\;
\vH^*_n=e^{a_s}\vH_n,\label{61}
\end{equation}
is physically senseless. This can be seen for example by making
again the following small thought experiment: Imagine we have the
following arbitrary but fixed mean velocity profile $\vH_1=\L
\vU\R$ based on some instantaneous (fluctuating) velocity field
$\vU$. Now, according to the left-hand side of \eqref{60} or
\eqref{61}, if we scale this mean profile $\vH_1$ by, say, a
constant factor $e^{a_s}=2$, we will get the two times amplified
mean velocity profile $\vH^*_1$. Hereby it should be noted that
this scaling is performed globally, i.e. for {\it all} points in
the considered physical space $\vx$ the mean velocity values
$\vH_1$ are scaled uniformly by a constant factor two.

Intuitively it's clear that a {\it globally} two times higher
amplitude in the mean profile can only go along with a {\it
globally} two times higher amplitude in the instantaneous
velocity. In other words, in order to account for a global scaling
$\vH^*_1=2\vH_1$ on the averaged level (the effect), the
underlying instantaneous velocity must transform accordingly
$\vU^*=2\vU$ on the fluctuating level (the cause), otherwise we
would not manage to reproduce this coherent amplification of a
factor two on the averaged level. $\phantom{\widetilde{\vU}}$

But now, if the instantaneous velocity $\vU$ globally scales (i.e.
for all points $\vx_n$ in physical space $\vx$) by a factor two,
then e.g. the two-point correlation function $\vH_2$ will globally
scale with a factor $e^{2a_s}=4$ as given in \eqref{60}, and {\it
not} as in \eqref{61} with the same factor $e^{a_s}=2$ as the mean
velocity $\vH_1$ is scaling.

\section{Proof No.2\label{SB}}
To prove that \eqref{1-24} in \cite{Oberlack14.1} is not admitted
as a symmetry transformation of the underlying LMN equations
\eqref{1-16}, we just have to show that the second summand of this
transformation \eqref{1-24}

\begin{align}
\widehat{f}_n  := & \;\: F(y_{(1)},\dotsc
,y_{(n)})\cdot\psi(\vv^\prime_{(1)})\cdot\delta^3(\vv^\prime_{(1)}-\vv^\prime_{(2)})
\cdots
\delta^3(\vv^\prime_{(1)}-\vv^\prime_{(n)})\label{141021:0932}\\[0.5em]
= &\;\: \lambda_{y_{(1)}}\cdots
\lambda_{y_{(n)}}\cdot\psi(\lambda_{y_{(1)}}v_{(1)1},v_{(1)2},v_{(1)3})
\nonumber\\
& \;\: \cdot
\delta(\lambda_{y_{(1)}}v_{(1)1}-\lambda_{y_{(2)}}v_{(2)1})\cdot\delta(v_{(1)2}-v_{(2)2})
\cdot\delta(v_{(1)3}-v_{(2)3})\nonumber\\
&\;\: \cdots
\delta(\lambda_{y_{(1)}}v_{(1)1}-\lambda_{y_{(n)}}v_{(n)1})\cdot
\delta(v_{(1)2}-v_{(n)2})\cdot
\delta(v_{(1)3}-v_{(n)3})\nonumber\\[0.5em]
= &\;\: \lambda_{y_{(1)}}\cdot
\psi(\lambda_{y_{(1)}}v_{(1)1},v_{(1)2},v_{(1)3})\nonumber\\
&\;\:
\cdot\delta\bigg(v_{(2)1}-\frac{\lambda_{y_{(1)}}}{\lambda_{y_{(2)}}}v_{(1)1}\bigg)
\cdot\delta(v_{(2)2}-v_{(1)2})
\cdot\delta(v_{(2)3}-v_{(1)3})\nonumber\\
&\;\:\cdots
\delta\bigg(v_{(n)1}-\frac{\lambda_{y_{(1)}}}{\lambda_{y_{(n)}}}v_{(1)1}\bigg)\cdot
\delta(v_{(n)2}-v_{(1)2})\cdot
\delta(v_{(n)3}-v_{(1)3}),\qquad\qquad\qquad\nonumber
\end{align}
where
\begin{equation}
\lambda_{y_{(k)}}:=\bigg(1-\frac{y^2_{(k)}}{H^2}\bigg)^{-1}\!
>0,\;\; k=1,\dotsc ,n,
\end{equation}

\noindent is not a solution to the considered equations
\eqref{1-16}. The reason is that since \eqref{1-16} is a linear
system and since the considered transformation \eqref{1-24} is a
coordinate-invariant superposition of two functions $f_n$ and
$\widehat{f}_n$ (\ref{141021:0932}), the transformed system
relative to the first function $f_n$ only stays then invariant if
the second superposed function $\widehat{f}_n$ (\ref{141021:0932})
forms a solution to this linear system \eqref{1-16}. Note that the
last identity in (\ref{141021:0932}) is based on the well-known
factorization and symmetry properties of the Dirac delta function.

Now, since the linear system of equations \eqref{1-16} consists of
three parts, the convective part on the left-hand side and the
pressure and viscous part on the right-hand side, and, since the
function $\widehat{f}_n$ (\ref{141021:0932}) to be inserted into
these three parts does {\it not} depend on the viscosity parameter
$\nu$, it is sufficient to show that if the viscous part of system
\eqref{1-16} is not vanishing then $\widehat{f}_n$
(\ref{141021:0932}) is not a solution to \eqref{1-16}, and hence
does not admit \eqref{1-24} as a symmetry transformation. To prove
this case, it's sufficient to show that this already happens on
the lowest level of order $n=1$.

\noindent For example, the viscous part of \eqref{1-16} gives the
result
\begin{gather}
 -\nabla_{\vv_{(1)}}\cdot
\left[\lim_{\vx_{(2)}\to\vx_{(1)}}\nu\Delta_{(2)}\int
d^3\vv_{(2)}\vv_{(2)}\widehat{f}_2\right]\hspace{4.5cm}\nonumber\\[0.5em]
=\;
-\partial_{v_{(1)1}}\left[-\frac{2\nu}{H^2}\cdot\lambda_{y_{(1)}}^2
v_{(1)1}\cdot\psi(\lambda_{y_{(1)}}v_{(1)1},v_{(1)2},v_{(1)3})\right]\nonumber\\[0.5em]
=\;
\frac{2\nu}{H^2}\lambda_{y_{(1)}}^2\left[\psi(\vv^\prime_{(1)})
+v^\prime_{(1)1}\cdot\frac{\partial}{\partial
v^\prime_{(1)1}}\psi(\vv^\prime_{(1)})\right]\neq
0,\qquad\label{141021:1121}
\end{gather}

\noindent which is non-zero, except if the function $\psi$ gets
restricted to satisfy the equation
\begin{equation}
\psi(\vv^\prime_{(1)})
+v^\prime_{(1)1}\cdot\frac{\partial}{\partial
v^\prime_{(1)1}}\psi(\vv^\prime_{(1)})=0, \label{141021:1137}
\end{equation}
for which then also the two remaining parts (the convective and
the pressure part) of \eqref{1-16} will vanish. Note that
$\vv^\prime_{(1)}=(v^\prime_{(1)1},v^\prime_{(1)2},v^\prime_{(1)3})$
is defined by \eqref{1-26} in having the components
\begin{equation}
v^\prime_{(1)i}=\delta_{1i}\cdot \lambda_{y_{(1)}}
v_{(1)1}+\delta_{i2}\cdot v_{(1)2}+\delta_{i3}\cdot v_{(1)3}.
\end{equation}
In other words, only if the function $\psi$ is restricted to the
solution of equation (\ref{141021:1137})
\begin{equation}
\psi(\vv^\prime_{(1)})=\frac{A(v^\prime_{(1)2},v^\prime_{(1)3})}{v^\prime_{(1)1}},
\label{141021:1304}
\end{equation}
where $A$ is some arbitrary integration function, then on the
lowest level of order $n=1$ the superposed function
$\widehat{f}_n$ (\ref{141021:0932}) is a solution of the linear
system \eqref{1-16}, and hence only then admits \eqref{1-24} as a
symmetry transformation; otherwise, in the general case without
specifying $\psi$ as \eqref{141021:1304}, transformation
\eqref{1-24} constitutes no symmetry.

For all higher orders beyond $n=1$, however, restriction
(\ref{141021:1137}) is only consistent to a weak (integral)
formulation, while in the strong (differential) formulation these
higher order restrictions force the function $\psi$ itself to
vanish, for which, thus, on the one side function $\widehat{f}_n$
(\ref{141021:0932}) turns into the trivial solution
$\widehat{f}_n=0$ of system \eqref{1-16}, and on the other side
where the symmetry transformation \eqref{1-24} then turns into the
trivial identity transformation~$f_n^*=f_n$.

Hence, in the strong formulation for the restrictive equations of
$\psi$ to all orders $n$, the transformation \eqref{1-24} is only
admitted by system \eqref{1-16} as a symmetry  if $\psi=0$. In the
weak formulation, however, transformation \eqref{1-24} is only
admitted as a symmetry if $\psi$ is restricted to the functional
structure (\ref{141021:1304}). But in this case we face the
unwanted problem that $\psi$ through (\ref{141021:1304}) then
exhibits a non-removable (unphysical) singularity at
$v^\prime_{(1)1}$, which definitely has not the smooth functional
form as suggested by the authors in~``Fig.~1"
(\cite{Oberlack14.1}, p.~8).

\bibliographystyle{jfm}
\bibliography{BibDaten}

\end{document}